\documentclass[sn-mathphys-num]{sn-jnl}

\usepackage{amsmath,amssymb,amsfonts}%
\usepackage{amsthm}%
\usepackage{mathrsfs}%
\usepackage[title]{appendix}%
\usepackage{xcolor}%
\usepackage{textcomp}%
\usepackage{manyfoot}%
\usepackage{booktabs}%
\usepackage{listings}%

\usepackage{adjustbox}
\usepackage{graphicx}%
\usepackage{multirow}%
\usepackage{makecell}
\usepackage{array}
\usepackage{multirow}

\usepackage{subfigure}
\usepackage{physics}
\usepackage{nomencl}
\makenomenclature

\usepackage{amsmath,amsfonts,bm}

\def\eqref#1{Eq.~(\ref{#1})}

\def\1{\bm{1}}

\DeclareMathAlphabet{\mathsfit}{\encodingdefault}{\sfdefault}{m}{sl}
\SetMathAlphabet{\mathsfit}{bold}{\encodingdefault}{\sfdefault}{bx}{n}

\def\gL{{\mathcal{L}}}

\newtheorem{definition}{Definition}
\newtheorem{theorem}{Theorem}

\newtheorem{example}{Example}

\usepackage{comment}
\usepackage{soul}
\usepackage{url}

\begin{document}

\title[Article Title]{Hybrid Quantum Neural Networks for Efficient Protein-Ligand Binding Affinity Prediction}

\author[1]{\fnm{Seon-Geun} \sur{Jeong}}\email{wjdtjsrms11@pusan.ac.kr}

\author[2]{\fnm{Kyeong-Hwan} \sur{Moon}}\email{drmoon@pusan.ac.kr}

\author*[1,2]{\fnm{Won-Joo} \sur{Hwang}}\email{wjhwang@pusan.ac.kr}

\affil[1]{\orgdiv{Department of Information Convergence Engineering}, \orgname{Pusan National University}, \orgaddress{\street{Busandaehak-ro 63beon-gil, Geumjeong-street}, \city{Busan}, \postcode{46241}, \country{Republic of Korea}}}

\affil[2]{\orgdiv{School of Computer Science and Engineering}, \orgname{Pusan National University}, \orgaddress{\street{Busandaehak-ro 63beon-gil, Geumjeong-street}, \city{Busan}, \postcode{46241}, \country{Republic of Korea}}}

\abstract{Protein-ligand binding affinity is critical in drug discovery, but experimentally determining it is time-consuming and expensive. Artificial intelligence (AI) has been used to predict binding affinity, significantly accelerating this process. However, the high-performance requirements and vast datasets involved in affinity prediction demand increasingly large AI models, requiring substantial computational resources and training time. Quantum machine learning has emerged as a promising solution to these challenges. In particular, hybrid quantum-classical models can reduce the number of parameters while maintaining or improving performance compared to classical counterparts.
Despite these advantages, challenges persist: why hybrid quantum models achieve these benefits, whether quantum neural networks (QNNs) can replace classical neural networks, and whether such models are feasible on noisy intermediate-scale quantum (NISQ) devices. This study addresses these challenges by proposing a hybrid quantum neural network (HQNN) that empirically demonstrates the capability to approximate non-linear functions in the latent feature space derived from classical embedding. The primary goal of this study is to achieve a parameter-efficient model in binding affinity prediction while ensuring feasibility on NISQ devices. Numerical results indicate that HQNN achieves comparable or superior performance and parameter efficiency compared to classical neural networks, underscoring its potential as a viable replacement.
This study highlights the potential of hybrid QML in computational drug discovery, offering insights into its applicability and advantages in addressing the computational challenges of protein-ligand binding affinity prediction.
}


\maketitle

\section{Introduction}
\label{sec1}

The effectiveness of a drug in interacting with a specific protein target is linked to the drug-target affinity, which is primarily determined by the structures of both the chemical and the protein~\cite{kitchen2004docking}. Binding of a molecule to a protein may start a biological process. This includes the activation or inhibition of an enzyme’s activity as well as the interaction between a drug molecule and its intended protein target. The binding is quantified by how strong the chemical compound. This metric quantifies how firmly the ligand, which is another term for the chemical compound, connects to the protein. Traditionally, dissociation ($K_d$), half inhibition concentrations ($IC_{50}$), and inhibition ($K_i$) constants have been utilized to represent experimentally determined binding affinities~\cite{zheng2019onionnet}. In drug discovery, a crucial selection criterion is the high binding affinity between a small molecule or short peptide and a receptor protein. Although the binding affinity could be measured directly through experimental methods, the time cost and financial expenses are extremely high due to insufficient known structures of protein-ligand complexes~\cite{burley2019rcsb, kitchen2004docking}. Therefore, protein-ligand binding affinity prediction can serves significant advantage in the drug discovery. 

In general, predicting the binding affinity can be categorized as: physics-based methods such as molecular docking~\cite{zhao2007flipdock}, molecular dynamics simulations~\cite{guterres2020improving}, have been widely used in binding affinity prediction. Similarity-based~\cite{brylinski2013nonlinear} or matrix factorization-based methods~\cite{cobanoglu2013predicting} predict binding affinity by utilizing the global similarity matrices of entire proteins or ligands.
However, these traditional structure-based methods still remain challenging to identifying the binding ligand from a large-scale chemical space through currently experimental methods. 
In recent years, artificial intelligence (AI) and machine learning (ML) models have emerged as a promising tool for binding affinity prediction. Leveraging large-scale protein-ligand datasets, these models have demonstrated superior accuracy by learning complex relationships between molecular structures and binding strength~\cite{stepniewska2018development}. For example, there are the deep learning methods for affinity prediction, such as Pafnucy~\cite{stepniewska2018development}, DeepAtom~\cite{rezaei2019improving}, TopologyNet~\cite{cang2017topologynet}, DeepDTA~\cite{zeng2021deep}, WideDTA~\cite{ozturk2019widedta}, and DeepDTAF~\cite{wang2021deepdtaf}. These models not only outperform traditional methods in prediction accuracy but also exhibit high-potential of drug discovery based on ML. Despite these advancements, challenges still remain. One major issue is the increasing size and complexity of ML models, which result in higher computational costs and longer training times\mbox{~\cite{askr2023deep}}.

Quantum machine learning (QML) combines the computational principles of quantum mechanics, such as superposition and entanglement, with ML algorithms, offering significant advantages in computation speed~\cite{moon2025qsegrnn, 10274707,JEONG2024608}. 
Recently, the universal approximation property (UAP) of quantum neural networks (QNNs) has been investigated, which is similar to the universal approximation theorem (UAT)~\cite{perez2020data} in AI theory. In particular, a multi-qubit QNN model defined a partial Fourier series can be a universal function approximator~\cite{PhysRevA.103.032430}. Although the expressivity of QNNs is strong, there is limitation of expressivity of QNNs even though the QNN is made sufficiently deep~\cite{wu2021expressivity}. 

One notable direction in QML research is the development of hybrid architectures that integrate classical ML and quantum computing. These hybrid models have demonstrated several advantages over purely classical counterparts~\cite{cao2018potential}. For instance, they often require fewer parameters, leading to faster computations while maintaining or even enhancing performance. This highlights the potential for QML to eventually replace classical ML in drug discovery, offering a transformative leap in efficiency and capability~\cite{blunt2022perspective}. However, hybrid quantum models are introduced but it remains challenges that the powerful expressivity comes from the classical part or the quantum part of hybrid models. Moreover, a systematic analysis of how parameters in the QNN affect the classes of functions that it can approximate is missing~\cite{yu2022power}.

Furthermore, it is essential to evaluate the feasibility of implementing a quantum model on noisy intermediate-scale quantum (NISQ) devices. NISQ devices face significant constraints, including a restricted number of quantum qubits, vulnerability to quantum computational errors, and short coherence times, which pose challenges to their implementation \cite{chen2023complexity, 10907925}. For example, the amplitude encoding method~\cite{sun2023asymptotically} requires deep circuit depth $O(poly(N))$, which may be infeasible due to the short coherence time of NISQ device.  In contrast, angle embedding maintains a constant circuit depth but requires $O(N)$ qubits~\cite{larose2020robust}. It requires a huge amount of qubits in large-dimensional input classical data. Therefore, designing a feasible encoding scheme on NISQ devices is crucial. 

\textbf{Contribution.} This study investigates and tackles several important challenges: 1) High computational costs and longer training times of classical ML models. 2)  The potential of QNN in place of classical neural networks (NNs). 
3) The feasibility of the proposed hybrid quantum model on NISQ devices. To address these challenges, we propose a novel QML-based method which is hybrid quantum DeepDTAF (HQDeepDTAF) to predict the protein-ligand binding affinity. In particular, to address the limitation of expressivity of QNN, the proposed model consists of a hybrid quantum model. To substitute the NN into hybrid quantum neural networks (HQNNs), we introduce data re-uploading models under the hybrid quantum model. 
For the target task, our structure is inspired by the DeepDTAF architecture, which consists of three separate modules: the entire protein module, the local pocket module, and the ligand simplified molecular input line entry system (SMILES) module. We follow the original module but the NN is substituted as a hybrid quantum model to achieve a parameter-efficient model in binding affinity prediction. Finally, we discuss the efficiency and feasibility of the proposed model. 
The main contributions of this study are summarized as follows: 
\begin{itemize}
    \item We propose a novel HQDeepDTAF for protein-ligand binding affinity prediction. Specifically, we introduce the hybrid embedding scheme to reduce the required qubit counts, and utilize classical regression network for prediction task.
    \item To address the limitation of expressivity of QNN, we propose a hybrid quantum framework for the target task. Specifically, our framework experimentally explores the various hybrid embedding schemes to find a suitable combination of quantum and classical. Moreover, to design the HQDeepDTAF model, we investigate the appropriate number of qubits and layers based on two key metrics: expressibility and entangling capability. 
    \item We evaluated the effectiveness of our proposed algorithm by conducting a comparative analysis with state-of-the-art benchmarks using the protein data bank bind (PDBbind) dataset~\cite{liu2017forging}. This assessment aimed to showcase the performance of our method in relation to existing state-of-the-art approaches. 
    \item We performed a noise simulation to evaluate the effectiveness of our proposed algorithm on NISQ devices. Furthermore, we discuss the efficiency and feasibility of the proposed model.
\end{itemize}

\section{Related Works}
This section presents the related work aligned with our
quantum algorithm. {\color{blue}}The related work includes the universal approximation of QNN, as well as classical and quantum models for protein-ligand binding affinity prediction.


\subsection{Universal Approximation of Quantum Neural Network}
The UAT establishes that deep NNs can approximate well-behaved functions with arbitrary accuracy, forming the basis of their expressive power. QNNs, as quantum analogues of classical NNs, have been similarly studied for their expressivity under the UAT framework.

Authors in \cite{cao2017quantumneuronelementarybuilding, torrontegui2019unitary, maronese2022quantum} have investigated the universal approximation of QNNs in terms of quantum activation functions or quantum neurons. They implemented the sigmoid and rectifier linear unit (ReLU) functions based on quantum neurons, which demonstrates that the quantum model can be used as a universal approximator. However, they required additional qubits, such as ancilla qubits, and multi-controlled gates for encoding the data into qubits.
\cite{yu2022power} has demonstrated that single-qubit QNNs can approximate any univariate function by mapping the model to a partial Fourier series. While the authors investigated the expressive power of a single-qubit QNN, this single-qubit QNN has a limited expressivity for multivariate functions. \cite{wu2021expressivity} demonstrated that sufficiently deep QNNs can approximate the target functions. While they exhibited that the loss is significantly reduced and approaches to zero, for adding nonlinearity in QNN, the quantum model requires additional ancillary qubits, which leads to serious computational resources.
\cite{PhysRevA.103.032430} has investigated the effect of data encoding on the expressivity of QNNs as function approximators. While the authors demonstrated that the multi-qubit QNNs have universality for multivariate functions, the quantum model requires exponential circuit depth, which is impractical to implement.

Authors in~\cite{inajetovic2023enabling} proposed variational QSplines and generalized QSplines to approximate non-linear activation functions in QNNs. While both methods achieve good approximation performance for specific functions like {sigmoid}, {ReLU}, and {ELU}, they are not general-purpose and require tailored formulations per activation function. In addition, these methods use amplitude encoding, which provides logarithmic qubit scaling with respect to input size but induces polynomially increasing circuit depth as data dimension grows. They also require additional ancilla qubits for inner product computation and spline evaluation, increasing hardware overhead on NISQ devices.
\cite{lubasch2020variational} proposed a variational quantum computing framework for solving nonlinear differential equations, notably the nonlinear Schr\"{o}dinger equation. Their approach introduces a quantum nonlinear processing unit which combines multiple copies of variational quantum states to handle nonlinearities, along with tensor network-inspired quantum circuits to implement efficient operators. While their method demonstrates the feasibility of quantum approaches for nonlinear problems, it relies on constructing a quantum circuit with polynomially increasing depth on qubits and data dimension, with precise gate structures and encoding of nonlinear terms via specially crafted quantum circuits with an ancilla qubit.

Unlike existing studies that focus solely on pure QNNs based on the UAT and require additional qubits and exponential circuit depth, we propose a parameter-efficient hybrid quantum model designed for practical implementation on NISQ devices. While we do not claim theoretical universality, we empirically demonstrate the capability of our HQNN to approximate nonlinear functions across multiple benchmark tasks. To evaluate its potential as a substitute for classical NNs, we conduct comparative experiments between HQNN, pure QNNs, and classical NNs using both univariate and multivariate function benchmarks. Furthermore, we examine the expressivity of HQDeepDTAF, which integrates HQNN, in the context of protein–ligand binding affinity prediction. The feasibility of the proposed model is assessed in terms of key NISQ constraints, including circuit depth and qubit count.

\subsection{Protein-Ligand Binding Affinity Prediction}
Leveraging large-scale protein-ligand datasets, ML models were developed for binding affinity prediction tasks. These models laid the foundation for modern advancements in drug discovery by utilizing protein-ligand interaction data to predict binding affinities with high accuracy.

Early approaches, such as Pafnucy~\cite{stepniewska2018development}, focused on deep learning-based regression techniques, incorporating features derived from protein-ligand structures to evaluate affinity. These efforts significantly improved the ability to predict drug-target interactions and were benchmarked rigorously against experimental datasets.
Further research was conducted, such as models like DeepDTA \cite{ozturk2018deepdta}, DeepDTAF \cite{wang2021deepdtaf}, DeepAtom \cite{li2019deepatom},  DeepAffinity \cite{karimi2019deepaffinity}, and WidtDTA \cite{ozturk2019widedta} which integrated convolutional neural networks (CNNs) for sequence-based feature extraction. These models emphasized the importance of representing both protein and ligand sequences effectively while demonstrating improved predictive capabilities over traditional docking methods. Additionally, TopologyNet \cite{cang2017topologynet} introduced a novel topology-based approach incorporating multi-task training to predict biomolecular properties alongside binding affinities.
To address the limitations of static representations, Zeng et al. \cite{zeng2021deep} employed multiple attention blocks to capture complex interaction patterns between ligands and binding sites. This attention mechanism not only enhanced the interpretability of predictions but also allowed the models to focus on critical interaction regions within the binding pockets. Similarly, hybrid frameworks, such as those described in Mohammad et al.'s research \cite{rezaei2019improving}, incorporated structural data and sequence-based information to refine affinity predictions further.

For the QML-based protein-binding affinity prediction, Domingo, Laia, et al.~\cite{domingo2023binding} proposed the hybrid quantum-classical 3D CNN. In particular, the flexible representation of quantum image (FRQI) method was used to encode the image data into quantum states. They reduced $20\%$ complexity than the classical counterparts while still maintaining optimal performance in the predictions. However, the FRQI method requires many quantum complex circuits for a single data sample. They showed the image in ($4\times4\times4$) blocks, $32832$ quantum circuits are required. The current NISQ devices have limitations in the available number of qubits and quantum gates as well as short decoherence time.
Dong, Lina, et al.~\cite{dong2023prediction} have shown that the quantum algorithm can achieve considerable accuracy, although the parameters used in the model have been remarkably reduced. In particular, quantum graph isomorphic networks, quantum graph convolution networks, and quantum CNNs have been constructed. They showed the potential of the hybrid quantum deep learning algorithm in bioinformatics. However, they left the noise effect on the proposed models in the NISQ devices.
Domingo, L., et al.~\cite{domingo2024hybridquantumclassicalfusionneural} substituted the end of the fully connected (FC) layer of both 3D CNN and spatial-graph CNN into a quantum fusion model. The study assessed the performance of the proposed model through a comprehensive comparison with a classical fusion benchmark. While the quantum fusion models outperformed their classical counterparts in both parameter efficiency and accuracy, the study did not address the practical feasibility of the quantum model or the limitations inherent to NISQ-era quantum devices.

Despite their critical impact on QML performance, previous studies have overlooked key factors such as the feasibility of the quantum model and NISQ device constraints such as qubit counts, decoherence, and noise. 
To address this gap, we propose a novel hybrid quantum model within the practical limits of NISQ-era quantum computing. Based on this approach, we address the major issue of classical ML models, which is the complexity of the model in terms of the number of parameters in this study.

\subsection{Hybrid Quantum Machine Learning}
Numerous studies have proposed a hybrid QML to leverage the advantage of quantum computing and classical ML algorithms in various domains.

\cite{tuysuz2021hybrid, JEONG2024608, 10274707} proposed a hybrid quantum-classical model for classification and prediction, demonstrating performance comparable to a classical model. They employed a combined FC layer with angle embedding or amplitude without a data re-uploading scheme for quantum data encoding. In addition, these studies did not consider noise effects in NISQ devices. \cite{senokosov2024quantum, ovalle2023quantum} introduced multiple parallel quantum circuits with angle embedding for image classification. This approach showed a reduction of computation time in terms of parameter count. The model outperformed a classical CNN or QCNN baseline on the MNIST dataset in accuracy. However, they didn't analyze noise effects experiments for multiple parallel quantum circuits for quantum features in NISQ devices.  
\cite{sinha2025nav} proposed an actor-critic-based quantum deep reinforcement learning model for collision-free navigation in self-driving cars, employing pure angle embedding in the critic. However, this embedding scheme is limited in handling high-dimensional inputs. While the proposed model demonstrated improved training stability and slightly higher average cumulative rewards compared to its classical counterpart, its performance significantly deteriorated under noise simulations.

While previous studies focus on pure quantum embeddings, FC-based hybrid embeddings, or multiple parallel quantum circuits, our approach adopts a hybrid architecture that differs in its embedding strategy. Specifically, we employ a data re-uploading scheme that integrates a classical embedding network with quantum angle embedding. Unlike prior work, our classical embedding can flexibly incorporate a wide range of encoders, such as NNs, CNNs, and other general-purpose architectures, enabling broader applicability and adaptability. This design supports flexible qubit allocation and enhances QNN performance. The proposed HQDeepDTAF is carefully tailored to NISQ constraints, offering a parameter-efficient architecture that ensures both performance and hardware feasibility.

\section{Background \& Preliminaries}
In this section, we present the necessary background to understand the proposed algorithm. We begin by describing the binding affinity prediction problem. Then, we introduce key concepts in QML, including quantum states, quantum encoding, and data re-uploading. Finally, we discuss the UAT, which forms the theoretical foundation of our approach.

\subsection{Machine Learning-based Binding Affinity Prediction}
\begin{figure*}[ht]%
\begin{center}
\includegraphics[width=\linewidth]{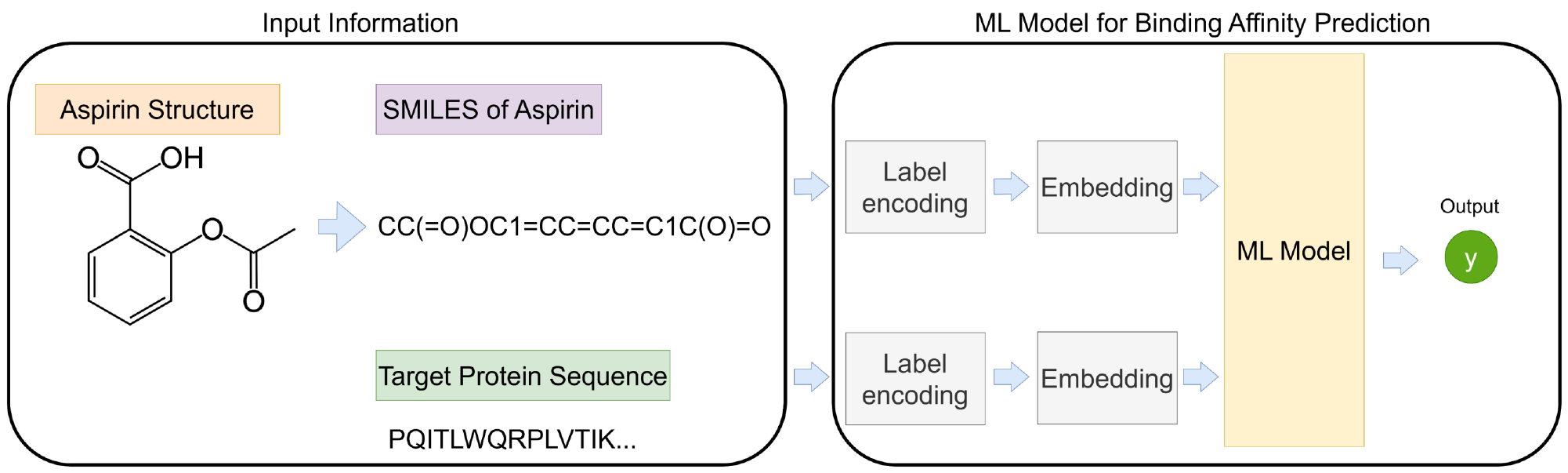}
\end{center}
\caption{Simple ML framework for binding affinity prediction. The input consists of the SMILES representation of Aspirin and an arbitrary target protein sequence. During model training, these inputs are encoded and fed into the ML model, which ultimately predicts the binding affinity value.}%
\label{fig:SimpleFramework}%
\end{figure*}
Binding affinity refers to the strength of the interaction between a drug and its receptor, and it is a key property for understanding how a drug’s structure influences its function~\cite{aloy2006structural}. Binding affinity prediction involves estimating the interaction strength between a ligand (e.g., a drug molecule) and its target protein. Traditionally, this has been measured through experimental methods, which are accurate but often time-consuming and costly~\cite{zheng2019onionnet}. To overcome these limitations, ML models have emerged as efficient alternatives~\cite{stepniewska2018development}.
ML-based binding affinity prediction typically relies on experimentally derived datasets, where binding affinities are expressed in terms of constants such as $K_d$, $IC_{50}$, and $K_i$. The objective of these models is to predict the binding strength based on molecular and protein information. Input representations can vary but commonly include the structures or sequences of proteins and ligands, which are often transformed into SMILES format.
As illustrated in Fig.~\ref{fig:SimpleFramework}, we present a simple ML framework for binding affinity prediction. In this example, the input consists of the SMILES representation of Aspirin and an arbitrary target protein sequence. These inputs are encoded and fed into the ML model, which is trained to predict the resulting binding affinity value.


\subsection{Quantum State}
The basic unit of information in quantum computation is one quantum bit, or qubit for short. A classical bit has a state based on Shannon information, with the state being either 0 or 1. However, the qubit can exist in a superposition of both the 1 and 0 states, the state of a single qubit is a unit vector in a $2$-dimensional Hilbert space $\mathbb{C}^2$, which is commonly denoted in Dirac notation as follows:
\begin{equation}\notag
\label{qubit}
|\psi\rangle = \alpha |0\rangle + \beta |1\rangle,
\end{equation}
where $\alpha$ and $\beta$ are complex numbers with $|\alpha|^2+|\beta|^2=1$. $\ket{0}=(1,0)^T$ and $\ket{1}=(0,1)^T$ are known as computational basis states. When qubits are measured, the result is always either 0 or 1; the probabilities of these outcomes depend on the quantum state of the qubits before the measurement. In addition, a quantum state of $n$ qubits can be represented as a normalized vector in the $n$-fold tensor product Hilbert space $\mathbb{C}^{2^n}$.

\subsection{Quantum Encoding}
Quantum encoding strategies, involving conversion classical data into quantum states, affects to the performance of quantum model directly~\cite{lloyd2020quantum}. 
Consequently, it is crucial to employ appropriate quantum embedding techniques to encode classical data into a quantum system. Numerous quantum embedding methods have been proposed~\cite{larose2020robust, sun2023asymptotically}, where angle embedding and amplitude embedding are widely used. An angle embedding method transforms the classical information into the angle of rotation $\theta \in [0, \pi]$. For example, given a input data $\mathbf{x}=[x_1,...,x_n]^T \in \mathbb{R}^N$, the angle embedding method maps all information into $O(N)$ qubits with a constant-depth quantum circuit as follows:  
\begin{equation}
U_{\phi}:\mathbf{x}\rightarrow |\phi(\mathbf{x})\rangle = \bigotimes_{i=1}^{N}\left(\cos\left(\frac{x_i}{2}\right)|0\rangle+\sin\left(\frac{x_i}{2}\right)|1\rangle\right),
\label{angleembedding}
\end{equation}
where $x_i \in [0,\pi)$ for all $i$. The angle embedding approach necessitates numerous qubits when dealing with substantial amounts of information. Thus, it is difficult to all information encodes to quantum states without loss of information due to the capability of current NISQ devices. 
In contrast, the amplitude embedding method has the advantage of reducing the number of qubits $O(\lceil \log(N) \rceil)$ for input data $\mathbf{x}\in \mathbb{R}^N$. Therefore, it can represent large amounts of information using a small number of qubits. For instance, for the normalized input data $\mathbf{\bar{x}}\in [0,1]^N$, each data $\bar{x}_i$ can be encoded into $\lceil \log(N) \rceil$ qubits as follows:
\begin{equation}
U_{\phi}:\mathbf{\bar{x}}\rightarrow|\phi(\mathbf{\bar{x}})\rangle =\sum_{i=1}^{2^{\lceil \log(N) \rceil}}\bar{x}_i|i\rangle,
\label{amplitude}
\end{equation}
where $\sum_i |\bar{x}_i|^2 = 1$. Although the amplitude embedding method requires a smaller number of qubits than the angle embedding method, the amplitude embedding 
has a circuit depth of $O(poly(N))$~\cite{sun2023asymptotically}. Therefore, the embedding method should be selected while considering the NISQ device capability and model performance.

\subsection{Data re-uploading}
The data re-uploading QNN model~\cite{perez2020data} is a generalized framework of QML models based on parameterized quantum circuits (PQCs). Given a set of input data $\mathbf{x}=\{x_1,..., x_n\}\in \mathbb{R}^N$ and a set of trainable parameters $\mathbf{\theta}=\{\mathbf{\theta_0},...,\mathbf{\theta_l}\}$, a data re-uploading QNN is a quantum circuit that consists of interleaved data encoding circuit blocks $S(\cdot)$ and PQC blocks $P(\cdot)$ as follows:
\begin{equation}
U_{\mathbf{\theta},L}=P(\theta_0)\prod_{i=1}^{L}S(\mathbf{x})P(\theta_i),
\label{Entangel}
\end{equation}
where $L$ denotes the number of PQC block layers. For the $N$ qubit system, the output of data re-uploading QNN model with observables $M_{\beta}(Z^{\beta_{1}}\otimes Z^{\beta_{2}} \otimes \cdots \otimes Z^{\beta_{n}})$ can be represented as follows:
\begin{equation}
f_{\mathbf{\theta},L}=\bra{0}^{\otimes N}U^{\dagger}_{\mathbf{\theta},L} M_{\mathbf{\beta}} U_{\mathbf{\theta},L}\ket{0}^{\otimes N}.
\label{outputdatareuploading}
\end{equation}

\subsection{Universal Approximation Theorem} 
The UAT plays an essential role in the development of NNs, which states that sufficiently wide or sufficiently deep defined NNs can approximate an arbitrary function with arbitrary accuracy~\cite{lu2017expressive}. This theorem lays the foundation of the expressive capability of NNs and serves as a basis for the success of NN applications. 

\begin{theorem}[Universal Approximation Theorem of classical NN~\cite{perez2020data}]
    For any Lebesgue integer function $f: \mathbb{R}^{N} \rightarrow \mathbb{R}$ and input data $\mathbf{x}=\{x_1,...,x_n\}\in \mathbb{R}^N$, there exists a FC classical NN with function $\varphi:\mathbb{R}\rightarrow \mathbb{R}$, the NN $F(\mathbf{x})=\sum_{i=1}^N\alpha_i \varphi(w_{i}x_{i} + b_{i})$ with $\alpha_{i}, b_{i} \in \mathbb{R}$ and $w_{i} \in \mathbb{R}^{N}$ such that $F$ with any precision $\varepsilon > 0$ satisfies   
    \begin{align}
        \left|f(\mathbf{x})-F(\mathbf{x})\right|dx < \varepsilon,
    \end{align}
\label{theorem:1}
\end{theorem}
where a lebesgue-integral function $f:\mathbb{R}^N \rightarrow \mathbb{R}$ is a Lebesque-measurable function satisfying $\int_{\mathbb{R}^N}|f(x)|dx<\infty$ which contains continuous functions. In classical NNs, $\varphi$ denotes the activation function, $w$ are the weights for each neuran, $b$ are the biases and $\alpha$ are the neuron weights that construct the output function. Therefore, this theorem establishes that it is possible to reconstruct any continuous function with a single layer NN of $N$ neurons. The proof of this theorem for ReLU activation function $\text{ReLU}(x) = \text{max}\{x,0\}$ can be found in~\cite{lu2017expressive}.

\section{Methodology}
\begin{figure*}[htp]%
\begin{center}
\includegraphics[width=\linewidth]{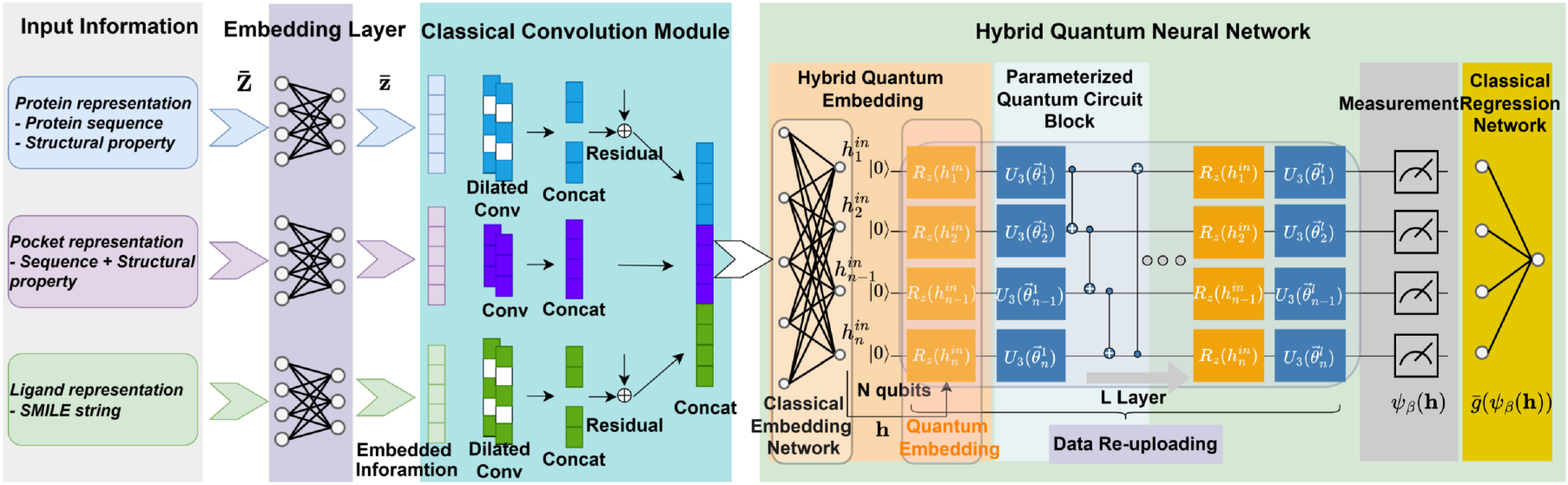}
\end{center}
\caption{Schematic of the HQDeepDTAF model. The representations of the protein, pocket, and ligand are provided as inputs. These inputs are first processed by an embedding layer. The classical convolution module consists of dedicated convolutional layers for each input type. The outputs from the three modules are then concatenated and passed to the HQNN. The hybrid quantum embedding, which combines a classical embedding network with a quantum embedding, encodes the data into a data re-uploading QNN. The measurement outcomes of the QNN are then passed to a classical regression network, which predicts the protein–ligand binding affinity.
}%
\label{fig:HQDeepDTAF}%
\end{figure*}

\subsection{Main architecture}
\label{sub:Main}
The proposed HQDeepDTAF is shown in Fig~\ref{fig:HQDeepDTAF}, consists of four main modules: 1) Input information; 2) Embedding layer; 3) Classical convolution module; 4) HQNNs for protein-ligand affinity prediction. For the input information, ligand representation, protein representation, and pocket representation are used as input data. These representations have proved to be beneficial for affinity prediction~\cite{wang2021deepdtaf}. Subsequently, input information is used for three different classical convolution modules for important feature extraction. The outcome of each module is concatenated to learn their interactions and predict the protein-ligand binding affinity via the HQNN. In the following, we first describe the input information and then describe classical convolution modules, and HQNN step by step. 


\subsubsection{Input Information}
As input data, we employed representations of the ligand, protein, and pocket, which can contribute to improved affinity prediction~\cite{wang2021deepdtaf}. Note that the input information of the model is 1-dimensional sequence data because the 3-dimensional structures of some proteins are still unknown. 
The details of the input information are listed as follows. 

\paragraph{Ligand representation}
The widely used 1-dimensional representation for ligand chemical structures is the SMILES \cite{weininger1988smiles}, which encodes molecular structures as character sequences that denote atoms, bonds, rings, and connectivity patterns. In this study, all ligand structure data format (SDF) files were converted into SMILES strings using Open Babel \cite{o2011open} to ensure a consistent molecular representation. Each SMILES string was subsequently mapped to a fixed-length numerical encoding scheme based on a predefined vocabulary of 64 distinct characters. Each character was assigned a unique integer identifier, such as ‘C’ represented as 42, ‘O’ as 48, ‘=’ as 40, ‘)’ as 31, and ‘(’ as 1. For example, the SMILES string “CC(=O)CC” was transformed into the sequence “42, 42, 1, 40, 48, 31, 42, 42”. 


\paragraph{Protein representation}
In this study, as global features, protein representation consists of sequence representation and structural property representation. These integrating sequence and structural information leads to more accurate and reliable prediction than using either representation alone~\cite{harding2024protein}.
Regarding protein sequence representation, a basic approach involves depicting the molecular structure as a one-dimensional sequence using a 20-amino acid alphabet.  Here, we utilized a one-hot vector consisting of 21 dimensions to encode the 21 various residue types present in protein sequences. Note that we opted for a 21-dimensional vector instead of a 20-dimensional one to accommodate non-standard residues present in certain proteins. The structural property representation encompassed both secondary structure elements (SSEs)~\cite{zhang2020probselect} and physicochemical attributes~\cite{wang2021deepdtaf}. To predict secondary structure for each sequence, the SSPro program~\cite{magnan2014sspro} was employed. Following this, SSEs were encoded using an 8-dimensional one-hot vector.  Physicochemical characteristics were encoded by an 11-dimensional vector. Consequently, a 19-dimensional vector was utilized to depict the structural property of each residue. In total, a 40-dimensional feature vector was employed for each residue to characterize global protein features, combining both sequence and structural property representations.

\paragraph{Pocket representation}
The pocket refers to a binding cavity in a protein, defined by specific physicochemical properties, shape, and location, which determine protein function. Protein-ligand interactions primarily depend on ligand binding to these pockets, which are composed of key amino acids from discontinuous sequences~\cite{wang2011ligand}. Therefore, a pocket representation is considered comprehensive for extracting local features. In protein-ligand binding affinity prediction, these local pocket features play a crucial role and are utilized as the primary input information. To encode local pocket features, a 40-dimensional feature vector was employed for each pocket residue. This vector combines sequence representation and structural property representation, as outlined in the protein representation.
\subsubsection{Embedding Layer}
We used an embedding layer to represent inputs with dense vectors in three modules. The embedding layer converts sparse, 1-dimensional sequence data inputs into dense vectors, making features more suitable for the model. In models where distinct features are processed through separate embedding layers, the layers are not shared because each feature represents fundamentally different information types. 
Separate embedding layers allow the model to learn feature-specific representations without interference.
\begin{definition}[Embedding Layer]
Given the input data $\bar{\mathbf{Z}}=[\bar{Z}_1,...,\bar{Z}_n]$, embedding layer transform input features into $K$-dimensional dense vectors as follows:
\begin{equation}
L_{e}:\bar{\mathbf{Z}} \rightarrow \bar{\mathbf{z}}=[\bar{z}_1,...,\bar{z}_k],
\end{equation}
\label{def:EmbeddingCNN}
\end{definition}
where the $k$-th component of the vector is given by $\sigma(w_{k1}\bar{Z}_1+\cdots+w_{kn}\bar{Z}_n)$, with $\sigma(\cdot)$ denoting the activation function and $w$ representing the trainable weight parameters.

\subsubsection{Classical Convolution Modules}
\begin{figure*}[htp]%
\begin{center}
\includegraphics[width=0.5\linewidth]{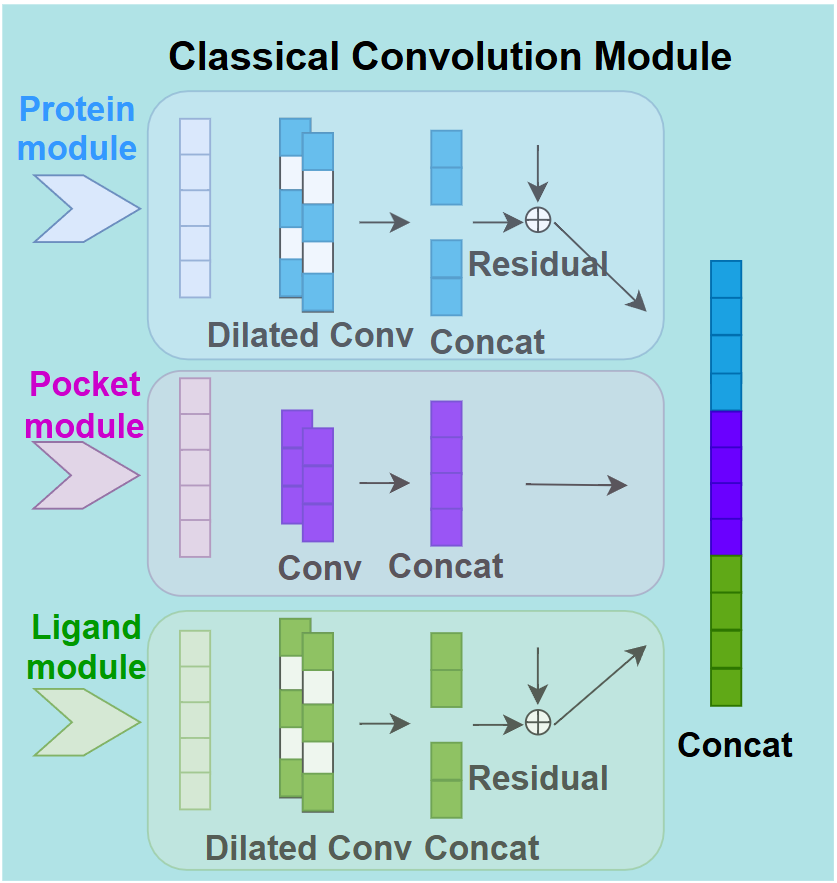}
\end{center}
\caption{Classical convolution modules, consisting of residual one-dimensional dilated convolutions in the protein and ligand modules, and a conventional one-dimensional convolution in the pocket module. The outputs from all three modules are concatenated.}%
\label{fig:ConvolutionModule}%
\end{figure*}
The embedded information obtained from Definition~\ref{def:EmbeddingCNN} is used to three different convolution modules as shown in Fig.~\ref{fig:ConvolutionModule}. With respect to the protein and ligand module, the dilated convolution~\cite{chen2019environmental} and residual learning~\cite{he2016deep} are introduced. For pocket module, we used vanilla convolution. For the design of classical convolution modules, we follows the DeepDTAF design~\cite{wang2021deepdtaf}. In the protein module, a one-dimensional dilated convolution with five different dilation rates was employed to account for long-range interactions in extended protein sequences. These dilated convolution layers were followed by a max pooling layer, similar to the ligand module. However, in the ligand module, the dilated convolution utilized four different dilation rates. For the pocket module, three one-dimensional conventional convolution layers were applied, progressively increasing the number of filters, followed by a max pooling layer. Finally, the outputs from the max pooling layers of all three modules were concatenated and passed into the HQNN component. 

\begin{definition}[Dilated Convolution~\cite{chen2019environmental}]
Given $F:Z^2 \rightarrow R$ is discrete function, $\Omega_r=[-r,r]^2 \cap Z^2$, and $k:\Omega_r \rightarrow R$ is a discrete filter of size $(2r+1)^2$. The dilated convolution operator $*_l$ with dilation rate $l$ is defined as follows:
\begin{equation}
(F*_l k) (p) = \sum_{s+lt=p}F(s)k(t),
\end{equation}
\label{def:DilatedConvolution}
\end{definition}
where $s$ and $t$ are subscripts of element vectors.  
To capture extensive interactions between protein features and ligand SMILES, Definition~\ref{def:DilatedConvolution} is employed, which expanded the effective size of the receptive field.

\subsubsection{Hybrid Quantum Neural Network}
\begin{figure*}[htp]%
\begin{center}
\includegraphics[width=\linewidth]{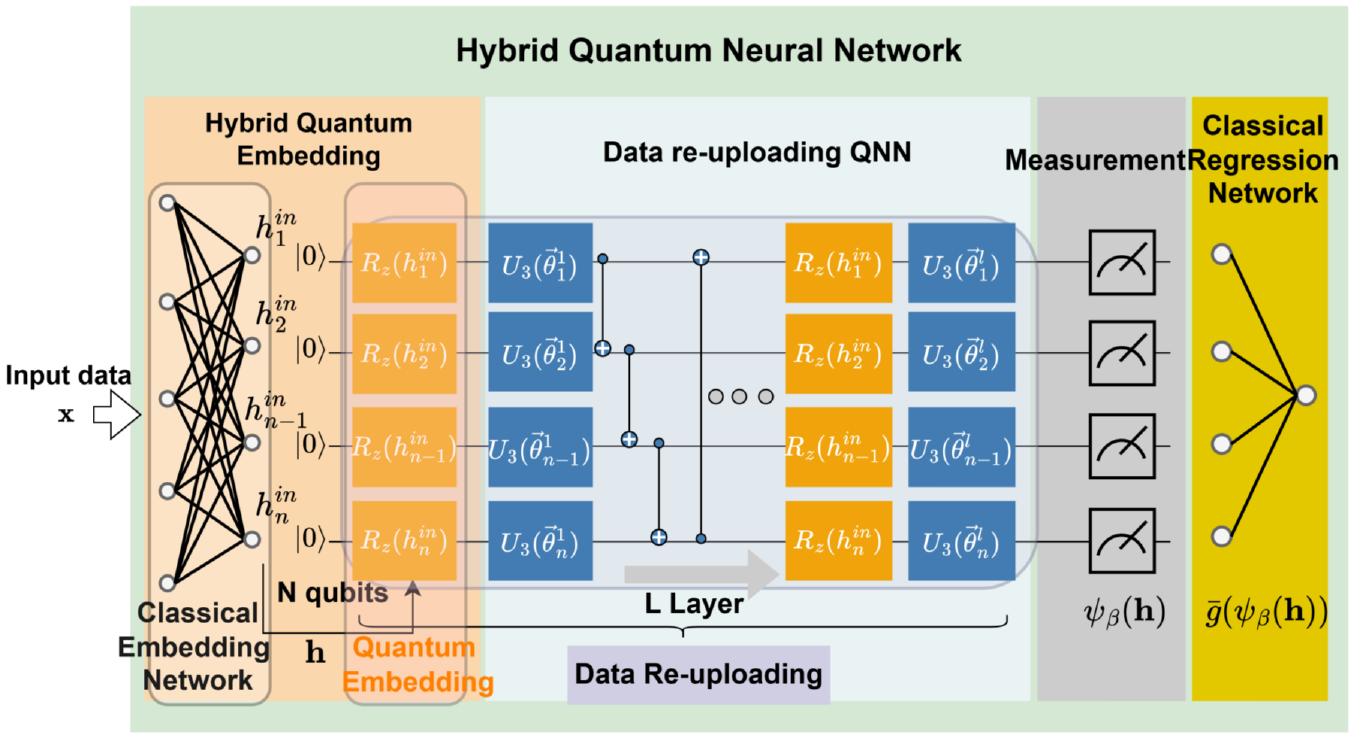}
\end{center}
\caption{Overview of the hybrid quantum neural network architecture. A classical embedding network generates the classical vector $\mathbf{h}$, which is encoded via quantum embedding into a data re-uploading QNN. The quantum output $\psi_{\beta}(\mathbf{h})$ is then passed to a classical regression network for final prediction.}%
\label{fig:HQNN}%
\end{figure*}

As shown in Fig.~\ref{fig:HQNN}, the proposed HQNN comprises a classical embedding network, a data re-uploading QNN, and a classical regression network. The hybrid quantum embedding, combining a classical embedding network with quantum embedding, allows flexible control over the number of qubits, which is essential for implementation on NISQ devices with limited qubit counts. Subsequently, the quantum computation for quantum states is conducted by the PQC blocks with data re-uploading scheme.

We use a classical embedding network, which can be NN, CNN, and various types of NN, as the embedding module to project input data $\mathbf{x} = \{x_1, ..., x_d\} \in \mathbb{R}^D$ into a lower-dimensional latent space $\mathbf{h} = [h_1^{in}, ..., h_n^{in}]^T$, where each component is given by:
\begin{equation}
h_n^{in} = \sigma(\omega_{n1}x_1 + \cdots + \omega_{nd}x_d),
\end{equation}
with weights $\omega$ and activation function $\sigma(\cdot)$. This latent vector $\mathbf{h}$ is then mapped to a Hilbert space using a quantum embedding $U_\phi$, as formalized below.

\begin{definition}[Hybrid Quantum Embedding]
Let $\mathcal{H}$ be a Hilbert space, and let $\mathbf{h}$ be the output of a classical embedding network. A quantum embedding circuit $U_\phi$ maps $\mathbf{h} \mapsto \ket{\phi(\mathbf{h})} \in \mathcal{H}$, where $\ket{\phi(\mathbf{h})}$ is the encoded feature vector.
\label{quantumembedding}
\end{definition}

\begin{example}[NN-Angle Embedding]
Using angle embedding, each $h_i^{in} \in \mathbf{h} \in \mathbb{R}^N$ is mapped as:
\begin{equation}
U_{\phi}(\mathbf{h}) = \bigotimes_{i=1}^{N}\left(\cos\left(\frac{h^{in}_i}{2}\right)\ket{0} + \sin\left(\frac{h^{in}_i}{2}\right)\ket{1}\right).
\end{equation}
\end{example}

We adopt a data re-uploading QNN architecture, in which the quantum circuit alternates between PQC blocks and data re-uploading blocks defined as
\begin{equation}
S(\mathbf{h})_H := e^{-i h_1^{in} H} \otimes \cdots \otimes e^{-i h_n^{in} H},
\end{equation}
where $H \in \{\sigma_x, \sigma_y, \sigma_z\}$ is a fixed Pauli operator. For simplicity, we employ angle embedding to encode classical data into quantum rotations.

The full QNN unitary with $L$ re-uploading layers is expressed as:
\begin{equation}
U(\mathbf{h}, \mathbf{\theta}) = P^{(L+1)} S(\mathbf{h})_H P^{(L)} \cdots S(\mathbf{h})_H P^{(1)} = P(\theta_0) \prod_{i=1}^{L} S(\mathbf{h})_H P(\theta_i).
\label{uni}
\end{equation}

The PQC block $P(\theta)$ uses single-qubit rotations and CNOT entanglers inspired by~\cite{schuld2020circuit}:

\begin{definition}[Parameterized Quantum Circuit Block] Assume $\forall i,~R_{i}\in \textbf{SU}(2)$ are single-qubit rotations with three trainable parameters of~\eqref{Req1}. $\forall j,~$ $T_{t_{j}}$ is a single-qubit unitaries, then the PQC blocks $P$ is defined as follows:
\begin{equation}
P=\prod_{j=0}^{n-1}E_{c_j}(T_{t_j})\prod_{i=0}^{n-1}R_{i}.
\label{Entangel}
\end{equation}
where the single-qubit unitary is defined as Pauli X. Thus, the $E_{c_j}(T_{t_j}) \in \text{U}(4)$ can be expressed as CNOT gate. In addition, $E_{c_j}(T_{t_j})$ are applied as a layer of $n/\text{gcd}(n,r)$ controlled gates, where $r$ is a hyperparameter called the range, gcd$(n,r)$ is the greatest common divisor of given $n$ and $r$, and $0<r<n$. Specifically, For $j\in[1,...,n/\text{gcd}(n,r)]$, the $j$-th 2-qubit gate $E_{c_j}(T_{t_j})$ of an PQC blocks have wire number $t_{j}=(jr-r)~\text{mod}~n$ as the control, wire number $c_{j}=jr~\text{mod}~n$ as target.
\label{def:PQCB}
\end{definition}
Each PQC block is hardware-compatible with NISQ constraints, consisting only of single-qubit rotations and CNOTs. As proven in Appendix~\ref{app1}, any such unitary can be decomposed using the Z-Y decomposition~\cite{nielsen2010quantum}.

\begin{definition}[Data Re-uploading QNN with Classical Embedding Network]
Given input $\mathbf{x} \in [0,1]^D$, embedding output $\mathbf{h}$, and observable $O_{\boldsymbol{\beta}} = Z^{\beta_1} \otimes \cdots \otimes Z^{\beta_n}$, the QNN output is defined as:
\begin{equation}
\psi_{\boldsymbol{\beta}}(\mathbf{h}) = \bra{0}^{\otimes N} U^{\dagger}(\mathbf{h}, \theta) O_{\boldsymbol{\beta}} U(\mathbf{h}, \theta) \ket{0}^{\otimes N}.
\label{quantummeasurement}
\end{equation}
\label{def:datare}
\end{definition}
Definition~\ref{def:datare} formalizes the data re-uploading QNN with a classical embedding network, distinguishing it from HQNN, which additionally includes a classical regression network. While increasing circuit depth improves QNN expressivity, pure QNNs still struggle to approximate general well-behaved functions due to inherent limitations~\cite{wu2021expressivity}. One remedy is to increase the Hilbert space dimension by using more qubits, but this is impractical on NISQ devices due to limited qubit counts and coherence times.
To overcome this, we incorporate a classical embedding network that both compresses input features and optimizes the quantum embedding during training, as supported by Theorem~\ref{theorem:1}. This allows for efficient use of limited qubits while maintaining expressive capacity.

To further enhance performance under NISQ constraints, we extend the structure defined in Definition~\ref{def:datare} by introducing a classical regression network, resulting in the HQNN architecture illustrated in Fig.~\ref{fig:HQNN}. The definition of the HQNN model is as follows.
\begin{definition}[Hybrid Quantum Neural Network]
Let $\mathbf{x} = \{x_1, ..., x_d\} \in [0,1]^D$ be the input data and $N \geq D$ the number of qubits. Let $\mathbf{h}$ be the output of a classical embedding network, and $O_{\boldsymbol{\beta}} = Z^{\beta_1} \otimes \cdots \otimes Z^{\beta_n}$ the observable with $\boldsymbol{\beta} \in \{0,1\}^N$. The QNN output $\psi_{\boldsymbol{\beta}}(\mathbf{h})$ follows Definition~\ref{def:datare}.
We define a classical regression network $\bar{g}:\mathbb{R}^K \rightarrow \mathbb{R}$ applied to $K$-dimensional QNN outputs $(K \leq N)$, where the final prediction is given by:
\begin{equation}
\bar{g}(\psi_{\boldsymbol{\beta}}(\mathbf{h})) = \sum_{i=1}^{K} \sigma(w_i \psi_{\beta_i} + b_i),
\end{equation}
with weights $w_i \in \mathbb{R}$ and biases $b_i \in \mathbb{R}$.
\label{def:HQNN}
\end{definition}

\subsection{Parameter Shift Rules for Quantum Neural Network Training} A quantum circuit driven by our framework learns a given task by updating parameters using parameter shift rules~\cite{mitarai2018quantum}.
Consider quantum states $\ket{\psi}$ comprising of $N$ qubits, unitary $U$ with trainable parameters $\mathbf{\theta}$, the expectation value with Pauli operators $B_j \in \{I,X,Y,Z\}^{\otimes n}$ can be represented as follows: $\langle \hat{\mathbf{B}}\rangle = \langle \psi|U^{\dagger}(\mathbf{\theta}) B_j U(\mathbf{\theta})|\psi\rangle$. Suppose expectation value is loss function $\gL (\theta)$ of a quantum circuit, then  
with a gradient-based strategy, the update of the trainable parameters $\theta$ of QNN at the step $t$ can be expressed as follows:
\begin{align}
    \theta_{t+1}=\theta_t - \eta \grad \gL (\theta_t),
\label{gradient}
\end{align}
where the gradient $\grad \gL (\theta_t)=\frac{1}{2}\langle \psi|U^{\dagger}(\mathbf{\theta}_t+\frac{\pi}{2}) B_j U(\mathbf{\theta}_t+\frac{\pi}{2})|\psi\rangle-\frac{1}{2}\langle \psi|U^{\dagger}(\mathbf{\theta}_t-\frac{\pi}{2}) B_j U(\mathbf{\theta}_t-\frac{\pi}{2})|\psi\rangle$. That is, the gradient can be calculated just by shifting $\pm \pi/2$ rotation on unitaries $U$. 

%

\section{Results}
In this section, we provide a detailed description and analysis of the experiments conducted to evaluate the performance of the proposed HQDeepDTAF. The subsections provide a brief overview of the dataset, experimental settings, and experiments employed in the prediction task. In addition, our experiment was designed with two primary objectives. The first objective was to demonstrate the function approximation capabilities of HQNN, which was analyzed to be more efficient in function approximation than QNN and NN. Subsequently, HQDeepDTAF experiments were conducted for protein-ligand binding affinity prediction. 

\subsection{Experimental Settings}
\begin{table*}[t]
    \small 
    \centering
    \caption{Summary of Models used for binding affinity prediction}
    \begin{adjustbox}{width=\textwidth,center} 
    \label{tab:BenchModel}
    \begin{tabular}{p{0.25\textwidth}p{0.65\textwidth}}
        \toprule
         \textbf{Model}& \textbf{Architecture Details}  \\
         \midrule
         DeepDTAF~\cite{wang2021deepdtaf} & For all embedding network, output dimension was $128$. For protein module, five 1D dilated convolutional layers with dilated rates of $1,2,4,8,16$. 
         For pocket module, three 1D convolutional layers with $32, 64, 128$ filters. 
         For ligand module, four 1D dilated convolutional layers with dilated rates of $1, 2, 4, 8$. 
         For all modules, convolutional filter size was 3 with max pooling layer after convolution. 
         The outcomes of each module are concatenated and input to the dense layers. 
         Dense consists of three layers with $128, 64, 1$ nodes, each followed by dropout layer of rate 0.5.  \\
         \midrule
         DeepDTA~\cite{ozturk2018deepdta} & The DeepDTA employs two CNN blocks followed by dense layer. Each CNN block consists of three 1D convolutions of 32, 64, 96 filters. Dense layer consists of 1024, 1024, and 512 neurons in sequence. \\
         \midrule
         Pafnucy~\cite{stepniewska2018development} & The Pafnucy architecture employs 3D convolutional layers with 64, 128, and 256 filters, each followed by max pooling. The output from the final convolutional layer is flattened and fed into dense layers consisting of 1000, 500, and 200 neurons in sequence. \\
         \midrule
         
         TopologyNet~\cite{cang2017topologynet} &$(\text{https://github.com/drmoon-1st/HQDeepDTAF}$) \\
         \midrule
         
         AEScore~\cite{aescore} &  
         ($\text{https://github.com/abdulsalam-bande/KDeep}$) \\
         \midrule

         $K_{deep}$~\cite{kdeep} & 
         ($\text{https://github.com/RMeli/aescore.git}$) \\
         \midrule

         HQDeepDTAF-Amplitude & For QNN with NN-based classical regression network, the concatenated results from each model are processed into $512$D data using zero padding. The $9$-qubits data re-uploading QNN followed. 
         \\
         \midrule
         HQDeepDTAF-NN-Amplitude & For HQNN with amplitude embedding, the NN-based classical embedding network with 512 nodes serves as input for a 9-qubit data re-uploading QNN consisting of 20 layers. 
         \\
         \midrule
         HQDeepDTAF-NN-Angle & For HQNN with angle embedding, the outcome of the NN-based classical embedding network with $9$ nodes serves as input for $9$-qubits data re-uploading QNN with $20$ layers. 
         \\
         \bottomrule
    \end{tabular}%
    \end{adjustbox}
\end{table*}


1) Data description: PDBbind database~\cite{liu2017forging} includes a collection of experimentally verified protein-ligand binding affinity expressed with $-\log(K_i),-\log(K_d),-\log(IC_{50})$ from the PDB~\cite{burley2019rcsb}. For our study, we selected the core $2016$ dataset from the PDBbind database version 2016, which is commonly utilized as a high-quality benchmark~\cite{wang2021deepdtaf}. This dataset encompasses a wide range of structures and binding information, making it suitable for assessing various docking techniques. 
The provided datasets included protein PDB files, pocket PDB files, and ligand SDF files, among others. We extracted the protein and pocket sequences from the PDB files. Additionally, we transformed the SDF files into SMILES strings.

2) Data settings: To create an effective representation format, it is essential to establish fixed lengths for protein sequences, pocket sequences, and SMILES strings, as they vary in length.
Based on \cite{wang2021deepdtaf}, to match the same experimental setting as the counterpart model on overall datasets, predetermined lengths for sequences of proteins, pockets, and SMILES strings were selected. 
As a result, we established fixed character limits for different sequence types: 1000 characters for protein sequences, 150 characters for SMILES strings, and 63 characters for pocket sequences. These limits were chosen to encompass approximately $90\%$ of the proteins, ligands, and pockets found in the datasets we analyzed. The sequences that are longer than the fixed characters were truncated and the sequences that are shorter than the fixed characters were 0 padded. The core 2016 set was compiled to test and evaluate our model. The Smith–Waterman similarity~\cite{zhao2013ssw} for each protein sequence in the core 2016 test set was at most $60\%$~\cite{zeng2021deep} to any sequence in the training set for $99\%$
of protein pairs. While training, we randomly load stochastic datasets for each iteration with random shuffling.

3) Baselines: In the experiments, to demonstrate the proposed HQDeepDTAF algorithm's effectiveness, we compared it with the state-of-the-art benchmarks, DeepDTAF~\cite{wang2021deepdtaf}, DeepDTA~\cite{ozturk2018deepdta}, Pafnucy~\cite{stepniewska2018development}, TopologyNet~\cite{cang2017topologynet}, AEScore \cite{aescore}, and $K_{deep}$ \cite{kdeep}. To ensure a fair and reproducible evaluation, we implemented all classical and quantum models using consistent training configurations and dataset preprocessing.
Both of DeepDTAF and HQDeepDTAF use the same structural classical convolution modules. In this study, we explored different embedding techniques to determine the most appropriate method for our proposed algorithm. Specifically, HQDeepDTAF-Amplitude incorporates a QNN based on amplitude embedding, coupled with a classical regression layer. In comparison, HQDeepDTAF-NN-Angle and HQDeepDTAF-NN-Amplitude adheres to Definition~\ref{def:HQNN}, utilizing a hybrid embedding approach as outlined in Definition~\ref{quantumembedding}. Note that, regarding HQDeepDTAF-NN-Amplitude model, at most $9$-qubits is introduced to set the similar total number of parameters with DeepDTAF. 
For baselines, we followed the architecture and training procedures described~\cite{wang2021deepdtaf, ozturk2018deepdta, stepniewska2018development, cang2017topologynet, aescore, kdeep}. However, TopologyNet was reimplemented using only protein sequence inputs, as its original implementation is not publicly available.
Table~\ref{tab:BenchModel} provides a comprehensive overview of the structural specifications for these models used in our experiments. All implementations, including reimplemented versions of some baselines, are made publicly available on our GitHub repository~{\url{https://github.com/drmoon-1st/HQDeepDTAF}}.

4) Training settings: 
In terms of convergence time, all models were trained for 20 epochs using a batch size of 16. The adaptive moment optimizer called AdamW~\cite{loshchilov2019decoupledweightdecayregularization} optimizer, which improves upon the original Adam by decoupling weight decay from the gradient update. In contrast to traditional L2 regularization, AdamW applies weight decay directly to the weights, rather than through the gradients. The update rule is given by:

\begin{equation}
\theta_{t+1} = \theta_t - \eta \left( \frac{\hat{m}_t}{\sqrt{\hat{v}_t} + \epsilon} + \lambda \theta_t \right)
\end{equation}
where $\theta_t$ denotes the model parameters at step $t$, $\hat{m}_t$ and $\hat{v}_t$ are the bias-corrected first and second moment estimates, $\eta$ is the learning rate, and $\lambda$ is the weight decay coefficient. This decoupled formulation leads to improved generalization and more stable training. In our experiment, we use a max $0.005$ learning rate, and the weight decay of $0.01$ was utilized for training each model. Mean squated error (MSE) is used for the loss function for training our model. MSE can be defined as follows:
\begin{equation}
MSE = \frac{1}{n} \sum_{i=1}^{n} \left( y_i - \hat{y}_i \right)^2
\end{equation}
Where $n$ denotes the number of samples, $y_i$ for the target that can be also defined as $-log(K_d/K_i)$, and $\hat y_i$ for the prediction value from our model.
The process was repeated five times, and the average prediction value was computed using the trained models with the lowest training loss value during the training process. The resulting values were used for comparison and discussion purposes. 
All experiments were conducted on a computer equipped with an AMD Ryzen 9 7950X 16-core Processor central processing unit (CPU) and 128 GB of RAM. We utilized PennyLane~\cite{bergholm2022pennylaneautomaticdifferentiationhybrid}, a widely used tool, to implement the quantum model. On the other hand, we employed PyTorch to implement the classical models. Note that, due to the limitations of our environment, we used only a classical computer in this study.

5) Metrics: 
In assessing protein-ligand binding affinity prediction, we compared the predicted values with experimentally determined affinity measurements. To gauge our model's effectiveness, we employed mean absolute error (MAE), and root mean square error (RMSE) as indicators of prediction accuracy. We aimed to evaluate the correlation between predicted and experimental affinity values using the Pearson correlation coefficient (R)~\cite{cohen2009pearson} and standard deviation (SD)~\cite{chesher2008evaluating} in regression analysis. The SD in regression was calculated as follows~\cite{wang2021deepdtaf}:
\begin{equation}\notag
SD = \sqrt{\frac{1}{N-1}\sum_{i=1}^N[y_i-(ap_i + b)]^2}
\end{equation}
Here, $N$ represents the total number of protein-ligand complexes, while $y_i$ and $p_i$ denote the actual and predicted affinity for the $i$th complex, respectively. The variables $a$ and $b$ correspond to the slope and intercept of the linear function between actual and predicted values.
Additionally, we utilized the concordance index (CI)~\cite{pahikkala2015toward}, which represents the probability that the predicted and true affinity values for two randomly chosen protein-ligand complexes are in a specific order. The equation for CI is expressed as~\cite{wang2021deepdtaf}:
\begin{equation}\notag
CI = \frac{1}{Z}\sum_{y_i > y_j}h(p_i - p_j),
\end{equation}
where $p_i$ represents the predicted value corresponding to the higher binding affinity value $y_i$, while $p_j$ denotes the predicted value for the smaller affinity value $y_j$. The variable $Z$ serves as a normalization constant, representing the total count of protein-ligand complexes. The function $h(u)$ is defined as follows: it equals 1.0 when $u>0$, 0.5 when $u=0$, and 0.0 when $u<0$. A stronger correlation between the model's predicted values and experimentally measured affinity values is indicated by a larger R value, lower SD, and higher CI value.

6) Experiments: We conducted a series of experiments to evaluate the performance of the proposed HQNN model. These include:
\begin{itemize}
    \item analysis of function approximation of HQNN
    \item analysis of layer and qubit counts effects
    \item analysis of the effect of the encoding scheme
    \item ablation study on HQNN components
    \item analysis of binding affinity prediction
    \item analysis of convergence speed
    \item analysis of the noise effects
\end{itemize}

\subsection{Analysis of Function Approximation of HQNN}
\begin{table*}[ht]
\centering
\caption{Experimental Dataset for Performance Evaluation}
\label{tdata}
\begin{adjustbox}{width=\textwidth,center} 
\begin{tabular}{ccccc}
\toprule
functions              & function type          &interval    & Training & Test   \\ \midrule
$\sin(5x)/(5x)$         & univariate function    &$x\in(0,3]$   & 200      & 100 \\
$\sin(5x_1)/(5x_1)+\sin(5x_2)/(5x_2)$          & multivariate function  &$x_1,x_2\in(0,3]$   & 200      & 100 \\ \botrule
\end{tabular}
\end{adjustbox}
\end{table*}

\begin{figure}[ht]
\subfigure[Univariate function for the function approximation experiments]{{\includegraphics[width=0.5\textwidth]{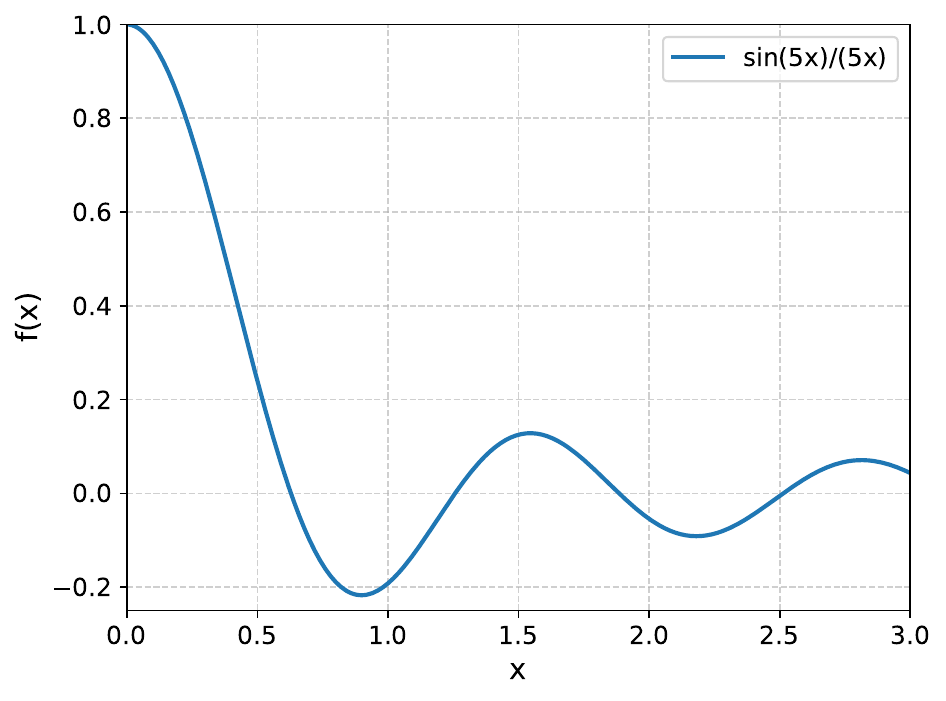} }}%
\label{fig.Targetfunctiona}
\vspace{0cm}
\subfigure[multivariate function for the function approximation experiments]{{\includegraphics[width=0.5\textwidth]{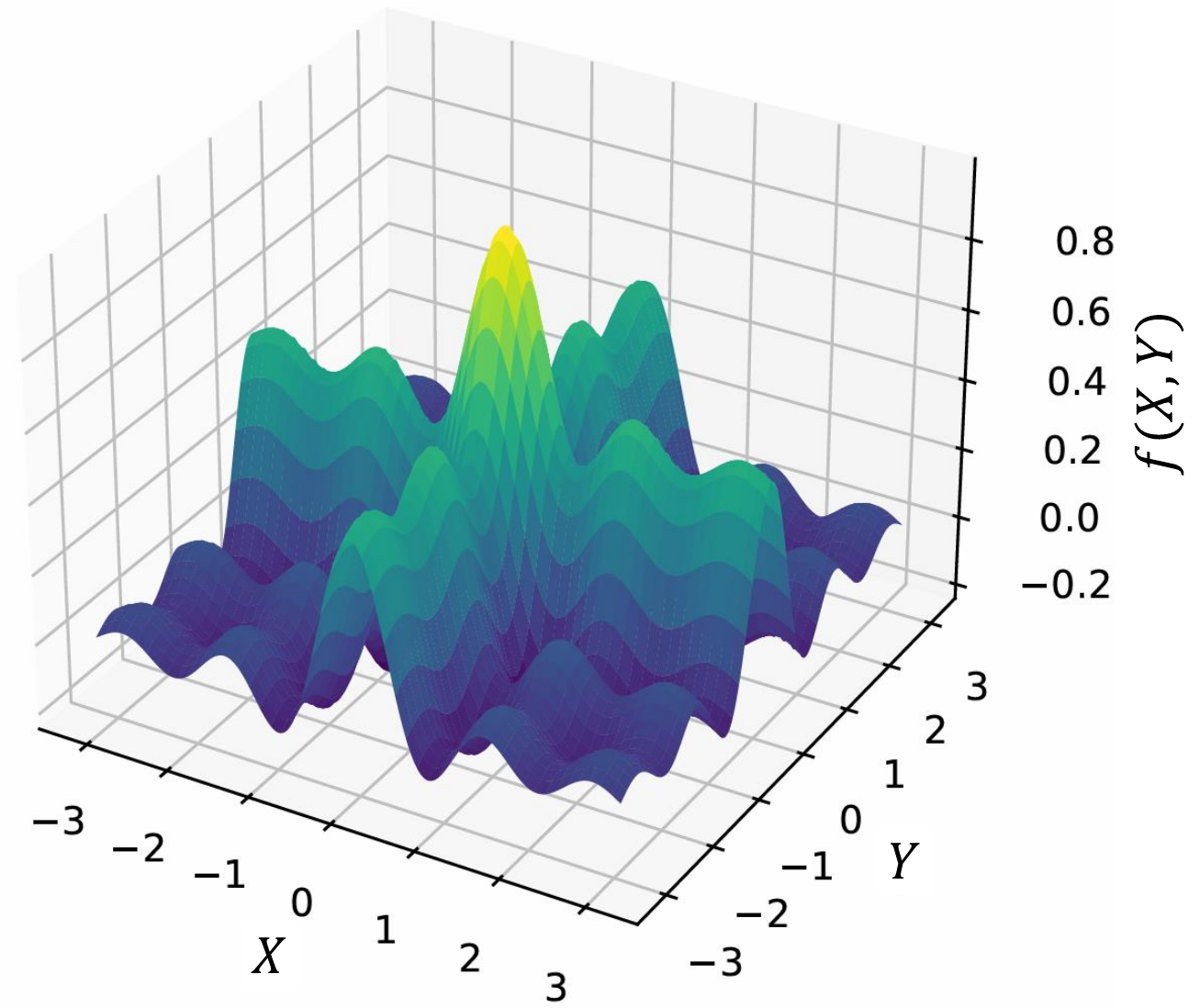} }}%
\label{fig.Targetfunctionb}
\caption{Target functions for the function approximation experiments}%
\label{fig.Targetfunction}%
\end{figure}
We investigate that HQNNs can effectively serve as non-linear function approximators and replace classical NNs. To demonstrate this, we concentrate on regression problems, as binding affinity prediction falls under this category of tasks. Herein, the experiments are widely used to demonstrate the capability of approximating non-linear functions of the model~\cite{wu2021expressivity, yu2022power, PhysRevA.103.032430}.
In our experimental setup, we evaluate both single-qubit and multi-qubit quantum models as function approximators for univariate and multivariate functions, respectively. For the single-qubit quantum model experiment, as shown in Fig.~\ref{fig.Targetfunction}~(a), we adpoted a damping function $f(x) = \sin(5x)/5x$ as target function. The dataset consists of $300$ data points uniformly sampled from the interval $x\in(0,3]$. For the multi-qubit quantum model experiment, as shown in Fig.~\ref{fig.Targetfunction}~(b), we adpoted a damping function $f(x) = \sin(5x_1)/5x_1+\sin(5x_2)/5x_2$ as target function. The dataset consists of $300$ data points uniformly sampled from the interval $x_1,x_2\in(0,3]$. The summarized information can be found in Table~\ref{tdata}. For each experiment, we compared NN, QNN, and HQNN. Regarding the NN, we fixed the number of nodes of each layer as 2 to set a similar number of parameters compared to quantum models. With respect to the HQNN, we investigate the effect of the classical embedding network in HQNN. Our findings allow us to contrast the performance of QNNs with NNs, focusing on the number of parameters required and the resulting errors. 

\begin{table*}[]
\centering
\caption{Univariate Function Approximation Performance of the NN, QNN, and HQNN}
\begin{adjustbox}{width=\textwidth,center} %
\begin{tabular}{cccccc}
\toprule
Model                             & \# Qubit  & Layer & \# C Params. & \# Q Params. & MSE                            \\ \midrule
\multirow{3}{*}{NN}               & \multirow{3}{*}{-}      & 1     & 2                   & -                 & $0.0794\pm0.0018  $ \\
                                  &       & 5     & 25                  & -                 & $0.0896\pm0.0370  $ \\  
                                  &      & 10    & 55                  & -                 & $0.1055\pm0.0298  $ \\ \midrule
\multirow{2}{*}{QNN}              & \multirow{2}{*}{1}      & 1     & -                   & 3                 & $0.0818\pm0.0355  $ \\
                                  &       & 5     & -                   & 15                & $0.0314\pm0.0477  $ \\ \midrule
\multirow{2}{*}{HQNN}             & \multirow{2}{*}{1}      & 1     & 4                   & 3                 & $0.0128\pm0.0145  $ \\
                                  &      & 5     & 4                   & 15                & $0.0002\pm0.0018  $ \\ \midrule
\multirow{2}{*}{HQNN without Cin} & \multirow{2}{*}{1}      & 1     & 2                   & 3                 & $0.0213\pm0.0149  $ \\
                                  &       & 5     & 2                   & 15                & $0.0003\pm0.0006  $ \\ \midrule
\multirow{3}{*}{NN}               & \multirow{3}{*}{-}      & 5     & 241                 & -                 & $0.0171\pm0.0319 $ \\
                                  &       & 5     & 865                 & -                 & $0.0090\pm0.0202  $ \\ 
                                  &       & 5     & 3265                & -                 & $0.0067\pm0.0148  $ \\ \botrule
\end{tabular}
\end{adjustbox}
\label{table:univariate}%
\end{table*}

\begin{table*}[]
\centering
\caption{Multivariate Function Approximation Performance of the NN, QNN, and HQNN}
\begin{adjustbox}{width=\textwidth,center} %
\begin{tabular}{cccccc}
\toprule
Model                               & \# Qubit  & Layer & \# C Params. & \# Q Params. & MSE                            \\ \midrule
\multirow{2}{*}{NN}                 &  \multirow{2}{*}{-}      & 1     & 3                   & -                 & $0.0669\pm0.0024  $ \\
                                    &       & 5     & 77                  & -                 & $0.0735\pm0.0191  $ \\ 
                                    &       & 10     & 57                  & -                 & $0.0754\pm0.0065  $ \\  \midrule                  
\multirow{2}{*}{QNN}                &  \multirow{2}{*}{2}      & 1     & -                   & 6                 & $0.0367\pm0.0174  $ \\
                                    &       & 5     & -                   & 30                & $0.0045\pm0.0056  $ \\ \midrule 
\multirow{2}{*}{HQNN}               &  \multirow{2}{*}{2}     & 1     & 9                   & 6                 & $0.0010\pm0.0011  $ \\
                                    &       & 5     & 9                   & 30                & $0.0005\pm0.0022  $ \\ \midrule 
\multirow{2}{*}{HQNN without Cin}   &  \multirow{2}{*}{2}     & 1     & 5                   & 6                 & $0.0049\pm0.0149  $ \\
                                    &       & 5     & 5                   & 30                & $0.0011\pm0.0014  $ \\ \midrule 
\multirow{3}{*}{NN}                 &  -     & 5     & 3297                & -                & $0.0144\pm0.0216 $ \\
                                    &  -     & 5     & 12737               & -               & $0.0124\pm0.0220  $ \\
                                    &  -     & 5     & 50049               & -                & $0.0104\pm0.0197  $ \\ \botrule
\end{tabular}
\end{adjustbox}
\label{table:multi}
\end{table*}

\begin{figure}[htp]
\subfigure[Univariate Function approximation result of the NN]{{\includegraphics[width=0.5\textwidth]{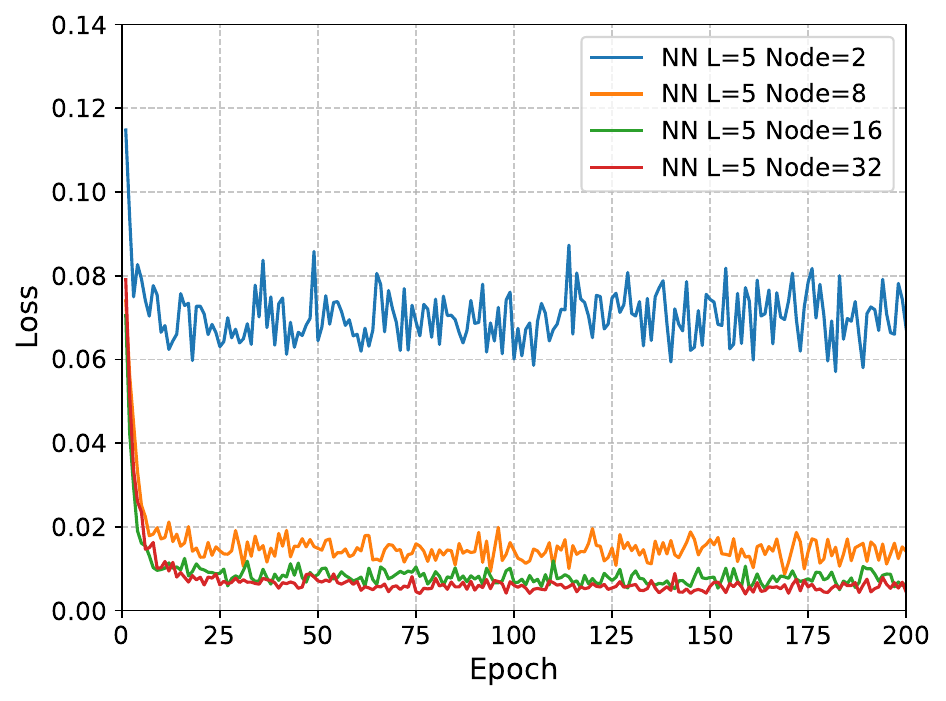} }}%
\label{fig.CarMNIST}
\vspace{0cm}
\subfigure[Univariate Function approximation result of the Quantum model]{{\includegraphics[width=0.5\textwidth]{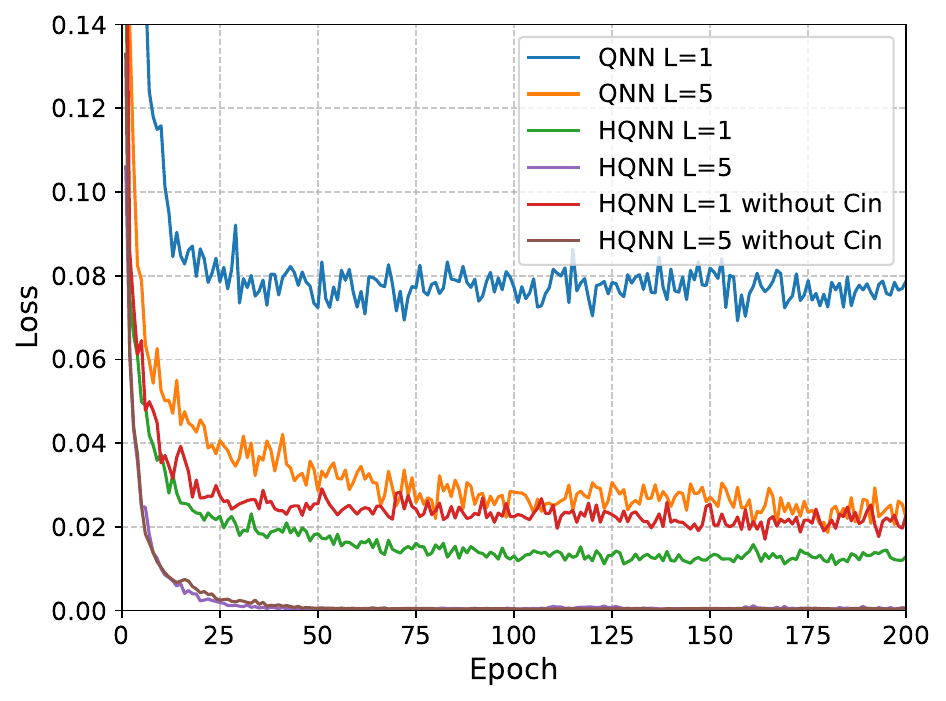} }}%
\label{fig.CarFashion}

\subfigure[Multivariate Function approximation result of the NN]{{\includegraphics[width=0.5\textwidth]{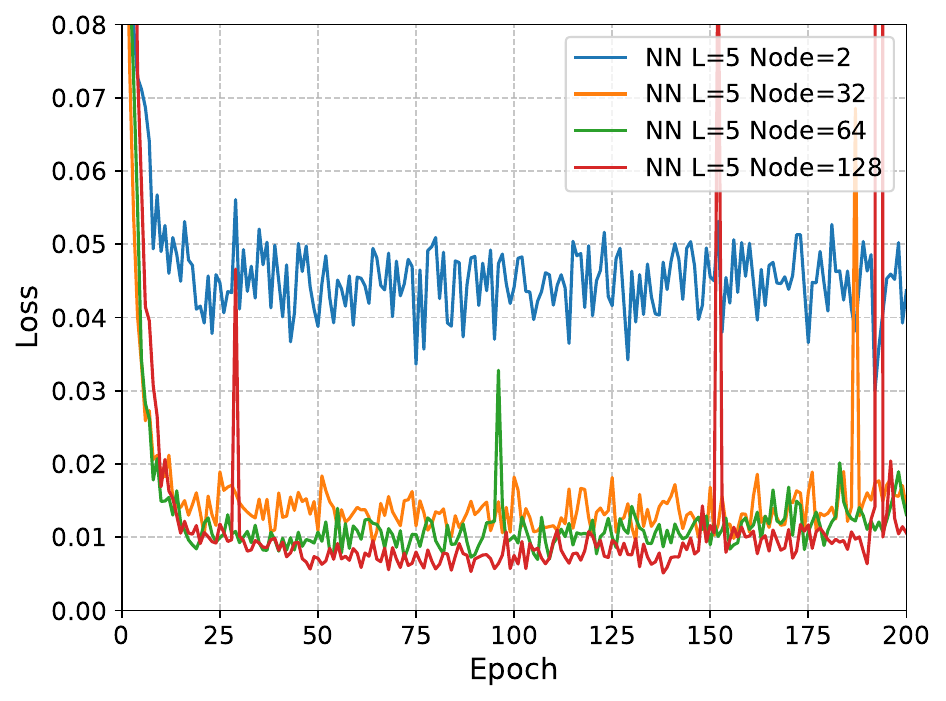} }}%
\label{fig.CarMNIST}
\vspace{0cm}
\subfigure[Multivariate Function approximation result of the Quantum model]{{\includegraphics[width=0.5\textwidth]{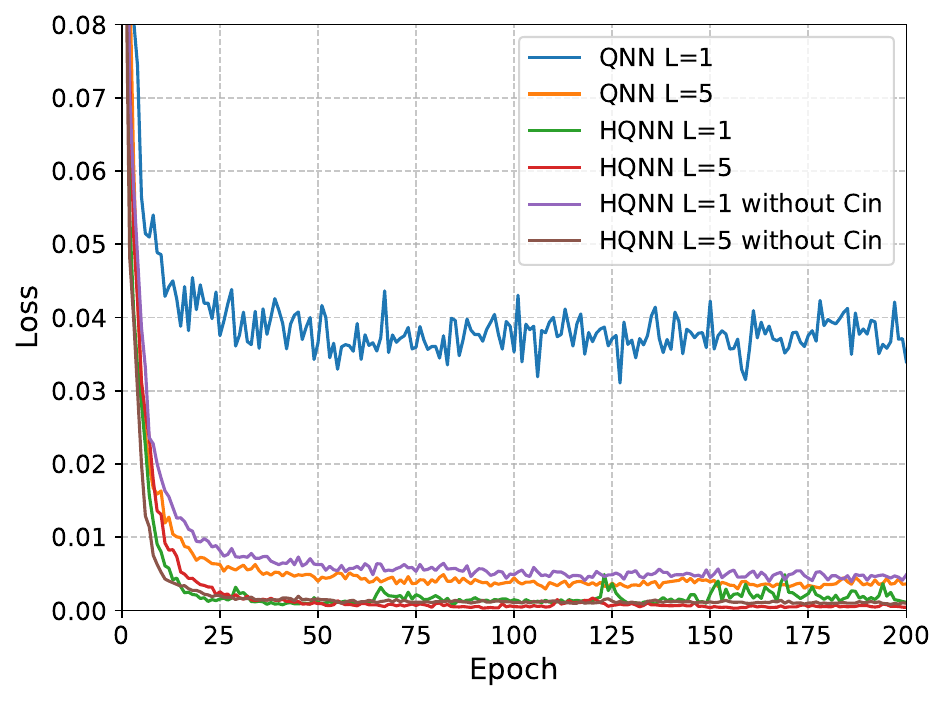} }}%
\label{fig.CarFashion}
\caption{Function approximation performance of the classical and quantum models for a simple function}%
\label{fig:singlefunction}%
\end{figure}

Table~\ref{table:univariate} displays the univariate function approximation performance of the NN, QNN, and HQNN, where $\#$ is the number of, C Params denote classical parameters, and Q Params means quantum parameters. For the NN, the performance decreases as the layer increases. In this process, we fixed the number of nodes of each layer as two. These results imply that it is critical to balance the dimensions of the network between width and depth. Note that these results are similar to model scaling~\cite{tan2019efficientnet}. 
To demonstrate these characteristics, we fixed the number of layers as 5 and only increased the number of nodes. As shown in Fig.~\ref{fig:singlefunction}~(a), the large number of nodes can obtain better performance, where $\text{Node}=2,8,16,32$ consists of $25, 241, 865, 3265$ classical parameters, respectively.  
In contrast, QNN has a single-qubit model, which means a fixed width. 
Despite the single-qubit model, the performance of the QNN is improved as the layer increases. 
Regarding HQNN, although HQNN with single layer has small classical parameters and quantum parameters, it can achieve better performance than NN comprising of $241$ parameters and QNN with 5 layers. In addition, compared with the HQNN with and without the classical embedding network with a single layer, we can verify that the synergy of the combination input NN and QNN is significant. In particular, HQNN can achieve dramatically high performance than QNN as the number of layers increases as shown in Fig.~\ref{fig:singlefunction}~(b). These results show that the hybrid quantum architecture might perform better with a smaller number of parameters than the NN and QNN for univariate function approximation.

Table~\ref{table:multi} exhibits the multivariate function approximation performance of the NN, QNN, and HQNN. For the NN, the performance decreases as the layer increases. As shown in Fig.~\ref{fig:singlefunction}~(c), we can verify that these results are the same as the univariate function approximation result of the NN, where $\text{Node}=2,32,64,128$ consists of $77, 3297, 12737, 50049$ classical parameters, respectively. Regarding the quantum model, hybrid models show higher performance than others, where HQNN outperforms others as shown in Fig.~\ref{fig:singlefunction}~(d). From these results, we can demonstrate the capability of HQNN for function approximation.  

\subsection{Analysis of Layer and Qubit Counts Effect}
To investigate the correlation between the layer and qubit counts and the performance of the PQC block generated by Definition~\ref{def:PQCB}, which is a component of the proposed HQNN, we introduce the expressibility and entangling capability~\cite{sim2019expressibility}. 
The expressibility and entangling capability metrics are commonly used to evaluate the performance of quantum circuits. Expressibility is known to be strongly correlated with trainability, indicating how effectively a quantum model can explore the parameter space. Meanwhile, high entangling capability reflects the circuit’s ability to efficiently represent complex quantum states, which is essential for solving high-dimensional tasks \cite{sim2019expressibility}. Therefore, quantum circuits with strong expressibility and high entangling capability are expected to exhibit superior performance. Based on these quantitative metrics, an appropriate number of qubits and circuit layers can be determined prior to applying the full hybrid architecture.

The expressibility of a quantum circuit refers to its capacity to generate quantum states that are uniformly distributed over the Hilbert space. Sim et al.~\cite{sim2019expressibility} proposed a quantitative measure of this property by comparing the distribution of state fidelities produced by a quantum circuit with the distribution of fidelities obtained from Haar-random states. Formally, expressibility is defined as:
\begin{equation}
\text{Expressibility}=D_{KL}\left(\hat{P}_{\text{QC}}(F(\mathbf{\Theta}))||P_{\text{Haar}}(F)\right),
\end{equation}
where $D_{KL}$ denotes the Kullback-Leibler (KL) divergence~\cite{kullback1951information}. The term $\hat{P}_{\text{QC}}(F(\mathbf{\Theta}))$ with parameter space $\Theta$ represents the estimated probability distribution of fidelities derived from sampling states generated by the quantum circuit, and $P_{\text{Haar}}(F)=(N-1)(1-F)^{N-2}$ is the analytical fidelity distribution corresponding to Haar-random states. Here, $F=|\langle \psi_{\theta}|\psi_{\phi}\rangle|^2$ denotes the fidelity between two quantum states, and $N$ is the dimension of the Hilbert space. 
Consequently, expressibility reaches zero when the two distributions match exactly. Thus, a lower value indicates that the circuit better approximates a Haar-random ensemble, implying greater expressivity and generalization potential.

The entangling capability of a quantum circuit is quantified by the average Meyer–Wallach entanglement~\cite{meyer2002global}:
\begin{equation}
\text{Entangling Capability}=\frac{1}{|S|}\sum_{\theta_{i}\in S}\mathcal{Q}\left(|\psi_{\theta_{i}}\rangle\right),
\end{equation}
where $\mathcal{Q}$ denotes the Meyer-Wallach entanglement measure, $S=\{\theta_{i}\}$ represents the set of sampled circuit parameters. A circuit that produces only separable (product) states yields an entangling capability of 0, while one that consistently generates highly entangled states approaches a value of 1.

\begin{figure}[ht]
\subfigure[Entangling capability results of QNN with $1, 5, 10, 15, 20$ layers and $2, 4, 7, 9$ qubits]{{\includegraphics[width=0.5\textwidth]{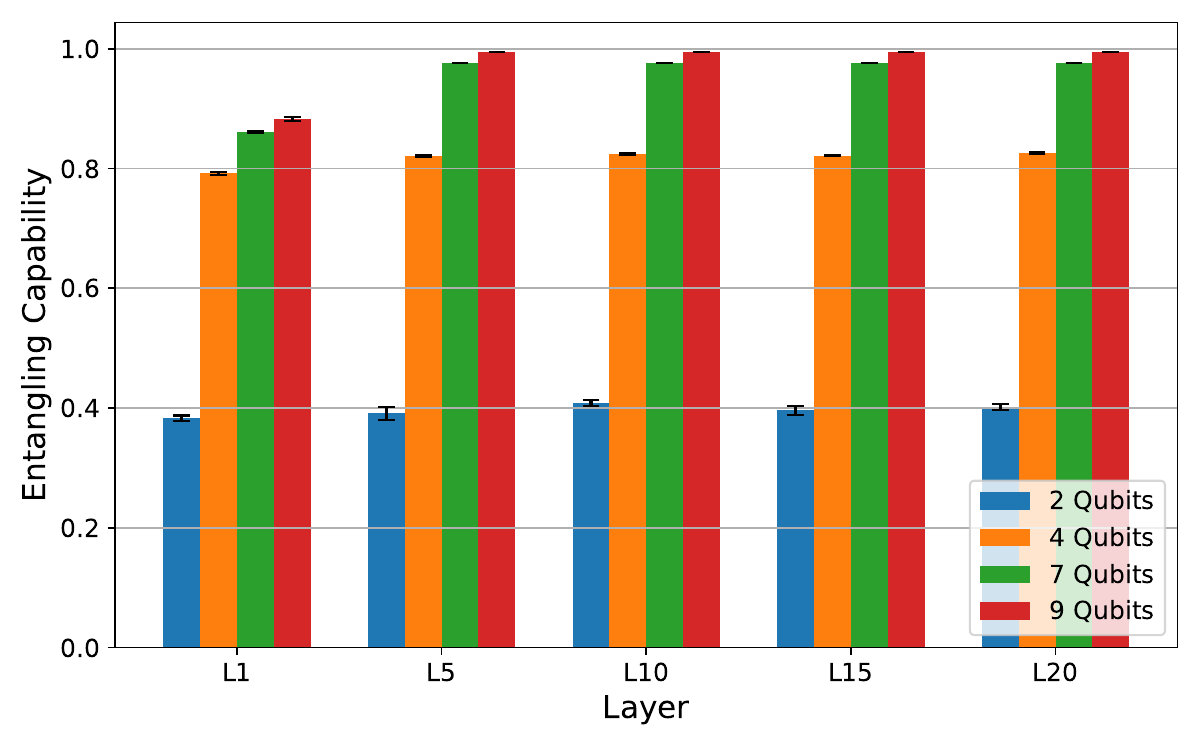} }}%
\label{fig.cq}
\vspace{0cm}
\subfigure[Expressivity results of QNN with $1, 5, 10, 15, 20$ layers and $2, 4, 7, 9$ qubits]{{\includegraphics[width=0.5\textwidth]{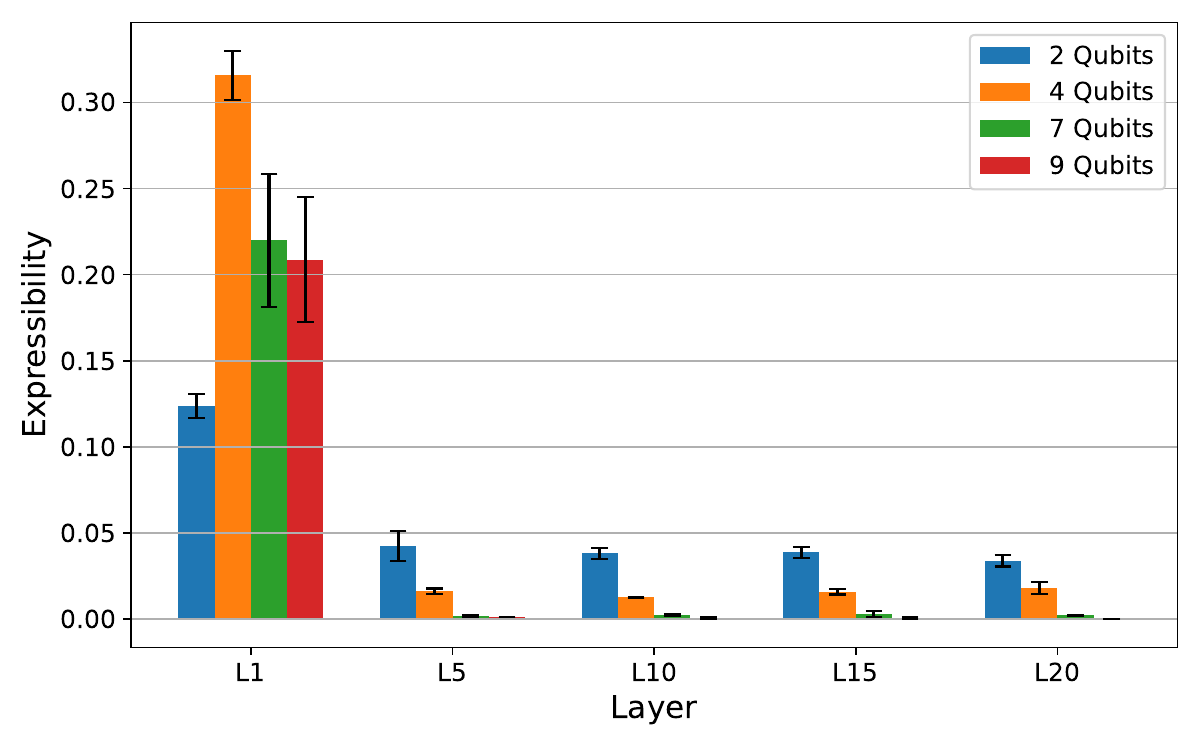} }}%
\label{fig.ct}
\caption{Correlation analysis of the PQC block with varying number of layers and qubits}%
\label{fig.enta}%
\end{figure} 

As shown in Fig.~\ref{fig.enta}, the entangling capability and expressibility of the PQC block are analyzed by varying the number of layers and qubits. Each configuration is evaluated over 4 independent runs using 1,000 shots per run, and the results are reported as the mean with standard deviation.
In Fig.~\ref{fig.enta}(a), the entangling capability increases consistently with the number of qubits across all layer settings. For a fixed number of qubits, deeper circuits (with more layers) also exhibit higher entangling capability. Notably, the PQC block with 9 qubits demonstrates significantly stronger entanglement compared to the 2-qubit configuration, approaching the theoretical maximum. Furthermore, configurations with more qubits and layers tend to show lower variance, indicating more stable entanglement generation.
Regarding the expressibility results shown in Fig.~\ref{fig.enta}(b), the expressibility of the PQC block increases with both the number of qubits and the number of layers. Notably, the PQC block with 9 qubits approaches maximum expressibility with a low standard deviation as the circuit depth increases.

These results provide a foundation for designing the HQNN with an appropriate number of qubits and layers. Given that the quantum circuit achieves near-maximal expressibility and entangling capability, the remaining performance of the HQNN largely depends on the choice of encoding scheme. Nevertheless, the number of qubits and circuit depth must be carefully selected to ensure the model’s feasibility on NISQ devices, where hardware limitations such as decoherence and qubit counts are significant constraints.

\subsection{Analysis of The Effect of The Encoding Scheme}

\begin{table*}[htbp]
\centering
\caption{Comparison of the Performance of HQDeepDTAF with varying embedding schemes, layer counts, and qubit counts.}
\label{tab:circuitdepth}
\begin{adjustbox}{width=\textwidth,center}
\begin{tabular}{llcccccc}
\toprule
\textbf{Embedding} & \textbf{} & \textbf{\# Qubit} & \textbf{Layer} & \textbf{Circuit Depth} & \textbf{\# C Param.} & \textbf{\# Q Param.} & \textbf{MAE} \\
\midrule
\multirow{5}{*}{Pure} & \multirow{5}{*}{Amplitude} & \multirow{5}{*}{9}
& 1  & $O(\mathrm{poly}(2^9)+12)$                & 96551 & 27 & 1.4018 $\pm$ 0.0500\\
& & & 5  & $O(\mathrm{poly}(2^9)\cdot5+60)$      & 96551 & 135 & 1.3835 $\pm$ 0.0464\\
& & & 10 & $O(\mathrm{poly}(2^9)\cdot10+120)$    & 96551 & 270 & 1.3977 $\pm$ 0.0246\\
& & & 15 & $O(\mathrm{poly}(2^9)\cdot15+180)$    & 96551 & 405 & 1.3199 $\pm$ 0.0639\\
& & & 20 & $O(\mathrm{poly}(2^9)\cdot20+240)$    & 96551 & 540 & 1.2984 $\pm$ 0.0958\\
\midrule
\multirow{20}{*}{Hybrid}
& \multirow{20}{*}{NN-Amplitude}
&  & 1  & $O(\mathrm{poly}(2^2)+4)$                 & 98084 & 6 & 1.1739 $\pm$ 0.0309\\
&   &   & 5  & $O(\mathrm{poly}(2^2)\cdot5+20)$     & 98084 & 30 & 1.1700 $\pm$ 0.0315\\
&   & \multicolumn{1}{c}{2}  & 10 & $O(\mathrm{poly}(2^2)\cdot10+40)$    & 98084 & 60 & 1.1642 $\pm$ 0.0245\\
&   &   & 15 & $O(\mathrm{poly}(2^2)\cdot15+60)$    & 98084 & 90 & 1.1747 $\pm$ 0.0057\\
&   &   & 20 & $O(\mathrm{poly}(2^2)\cdot20+80)$    & 98084 & 120 & 1.1684 $\pm$ 0.0093\\
\cmidrule{3-8}
& &  & 1  & $O(\mathrm{poly}(2^4)+6)$        & 102706 & 12 & 1.1743 $\pm$ 0.0408\\
&   &   & 5  & $O(\mathrm{poly}(2^4)\cdot5+30)$     & 102706 & 60 & 1.1704 $\pm$ 0.0217\\
&   & 4  & 10 & $O(\mathrm{poly}(2^4)\cdot10+60)$    & 102706 & 120 & 1.1647 $\pm$ 0.0179\\
&   &   & 15 & $O(\mathrm{poly}(2^4)\cdot15+90)$    & 102706 & 180 & 1.1654 $\pm$ 0.0150\\
&   &   & 20 & $O(\mathrm{poly}(2^4)\cdot20+120)$   & 102706 & 240 & 1.1623 $\pm$ 0.0514\\
\cmidrule{3-8}
& &  & 1  & $O(\mathrm{poly}(2^7)+10)$        & 145829 & 21 & 1.1937 $\pm$ 0.0241 \\
&   &   & 5  & $O(\mathrm{poly}(2^7)\cdot5+50)$     & 145829 & 105 & 1.1771 $\pm$ 0.0234 \\
&   & 7  & 10 & $O(\mathrm{poly}(2^7)\cdot10+100)$   & 145829 & 210 & 1.1684 $\pm$ 0.0217 \\
&   &   & 15 & $O(\mathrm{poly}(2^7)\cdot15+150)$   & 145829 & 315 & 1.1427 $\pm$ 0.0092\\
&   &   & 20 & $O(\mathrm{poly}(2^7)\cdot20+200)$   & 145829 & 510 & 1.1041 $\pm$ 0.0188 \\
\cmidrule{3-8}
& &  & 1  & $O(\mathrm{poly}(2^9)+12)$        & 145831 & 27 & 1.1859 $\pm$ 0.0396 \\
&   &   & 5  & $O(\mathrm{poly}(2^9)\cdot5+60)$     & 145831 & 135 & 1.1624 $\pm$ 0.0194 \\
&   & 9  & 10 & $O(\mathrm{poly}(2^9)\cdot10+120)$   & 145831 & 270 & 1.1432 $\pm$ 0.0099 \\
&   &   & 15 & $O(\mathrm{poly}(2^9)\cdot15+180)$   & 145831 & 405 & 1.1350 $\pm$ 0.0109 \\
&   &   & 20 & $O(\mathrm{poly}(2^9)\cdot20+240)$   & 145831 & 540 & 1.0856 $\pm$ 0.0216 \\
\midrule
\multirow{20}{*}{Hybrid} & \multirow{20}{*}{NN-Angle}
&  & 1  & $O(3)$             & 97314 & 6 & 1.1687 $\pm$ 0.0497\\
&   &   & 5  & $O(15)$       & 97314 & 30 & 1.1716 $\pm$ 0.2515\\
&   & 2  & 10 & $O(30)$      & 97314 & 60 & 1.1652 $\pm$ 0.0067\\
&   &   & 15 & $O(45)$       & 97314 & 90 & 1.1658 $\pm$ 0.0783\\
&   &   & 20 & $O(60)$       & 97314 & 120 & 1.1669 $\pm$ 0.0387\\
\cmidrule{3-8}
& &  & 1  & $O(5)$           & 98086 & 12 &  1.1724 $\pm$ 0.0088\\
&   &   & 5  & $O(25)$       & 98086 & 60 &  1.1555 $\pm$ 0.0277\\
&   & 4  & 10 & $O(50)$      & 98086 & 120 &  1.1488 $\pm$ 0.0278\\
&   &   & 15 & $O(75)$       & 98086 & 180 &   1.1337 $\pm$ 0.0371\\
&   &   & 20 & $O(100)$      & 98086 & 240 &   1.1302 $\pm$ 0.0411\\
\cmidrule{3-8}
& &  & 1  & $O(11)$          & 99244 & 21 & 1.1890 $\pm$ 0.0305\\
&   &   & 5  & $O(55)$       & 99244 & 105 & 1.1641 $\pm$ 0.0224\\
&   & 7  & 10 & $O(110)$     & 99244 & 210 & 1.1404 $\pm$ 0.0184\\
&   &   & 15 & $O(165)$      & 99244 & 315 & 1.1150 $\pm$ 0.0190\\
&   &   & 20 & $O(220)$      & 99244 & 420 & 1.0946 $\pm$ 0.0133\\
\cmidrule{3-8}
& &  & 1  & $O(13)$          & 100016 & 27 & 1.1794 $\pm$ 0.0330\\
&   &   & 5  & $O(11)$       & 100016 & 135 & 1.1651 $\pm$ 0.0159\\
&   & 9  & 10 & $O(55)$      & 100016 & 270 & 1.1350 $\pm$ 0.0220\\
&   &   & 15 & $O(195)$      & 100016 & 405 & 1.1138 $\pm$ 0.0154\\
&   &   & 20 & $O(260)$      & 100016 & 540 & 1.0821 $\pm$ 0.0049\\
\bottomrule
\end{tabular}
\end{adjustbox}
\end{table*}

To evaluate the effectiveness of different encoding strategies in our HQDeepDTAF model for protein–ligand binding affinity prediction, we compare prediction accuracy across various configurations involving different embedding types (pure amplitude, NN-Amplitude, and NN-Angle), qubit counts, and circuit depths. 

Table~\ref{tab:circuitdepth} summarizes the performance across these settings. The HQDeepDTAF model receives a 384-dimensional input vector, requiring 9 qubits when using pure amplitude embedding. Hence, the pure amplitude model is evaluated at 9 qubits across increasing circuit depths (layers 1 to 20). As shown, both increased qubit counts and deeper circuits generally lead to better performance across all embedding types. However, the pure amplitude embedding consistently underperforms compared to its hybrid counterparts, especially at lower depths.
From a circuit complexity perspective, both the pure amplitude and NN-Amplitude embeddings share a similar depth, formulated as $O(\mathrm{poly}(2^n)L + L(3 + n))$,
where $n$ is the number of qubits and $L$ is the circuit depth. This overhead arises from the quantum embedding scheme used in both methods. In contrast, NN-Angle embedding exhibits significantly shallower circuit depth, approximated as
$O(L + L(3 + n))$, as it does not require exponentially scaling amplitude encoding.
Moreover, NN-Angle embedding yields comparable or even superior performance to NN-Amplitude while using fewer classical parameters. For instance, the 9-qubit NN-Angle model with 20 layers achieves an MAE of 1.0821, outperforming NN-Amplitude (1.0856) and significantly outperforming the pure amplitude embedding (1.2984) under identical qubit and layer conditions.

These findings confirm that NN-Angle embedding with 9 qubits is a particularly efficient and scalable configuration for HQDeepDTAF, offering a favorable trade-off among circuit depth, parameter efficiency, and predictive accuracy, making it well-suited for NISQ devices. In addition, from the perspective of the UAT, the key distinction between pure amplitude embedding and hybrid approaches (NN-Amplitude and NN-Angle) is the presence of a classical embedding network. While pure amplitude directly encodes the input into a quantum state, it lacks a trainable mechanism to adapt the input representation. In contrast, NN-based embeddings use a classical network to transform the input into a task-specific latent space before quantum encoding. This added expressivity enhances the model’s approximation properties, as reflected in the consistently lower MAEs of hybrid methods. For example, at layer 1, NN-Amplitude achieves 1.1859 vs. 1.4018 for pure amplitude. As layers deepen, this performance gap persists, underscoring the classical embedding’s essential role in enabling HQNNs to effectively approximate complex functions.

\subsection{Ablation Study on HQNN Components}

\begin{table*}[h]
\caption{Comparison of prediction performance and parameter allocation across NN, HQNN with frozen quantum layers, and HQNN with trainable quantum layers}
    \centering
    \begin{adjustbox}{width=\textwidth,center}
\begin{tabular}{llllll}
\hline
Models                  & \# Qubit & \# C Param. & \# Q Param. & Layer & MAE                            \\ \hline
Classical NN            & -        & 101336      & -           & 20    & 1.205 $\pm$ 0.202 \\
NN-Amplitude (Freezing) & 9        & 145831      & 540         & 20    & 1.167 $\pm$ 0.033 \\
NN-Amplitude (Training) & 9        & 145831      & 540         & 20    & 1.086 $\pm$ 0.039 \\
NN-Angle (Freezing)     & 9        & 100016      & 540         & 20    & 1.191 $\pm$ 0.012 \\
NN-Angle (Training)     & 9        & 100016      & 540         & 20    & 1.082 $\pm$ 0.004 \\ \hline
\end{tabular}
\end{adjustbox}
\label{tab:ablation}
\end{table*}
To evaluate the effectiveness of the quantum component in the proposed HQDeepDTAF architecture, we conducted an ablation study across five configurations, as presented in Table~\ref{tab:ablation}. First, we used a fully classical NN (denoted as Classical NN) as a baseline. Then, we examined two variants of the HQNN architecture based on different quantum embedding methods: 1) NN-Amplitude and 2) NN-Angle. For each embedding type, we considered two settings: one with frozen quantum parameters (Freezing), and the other with trainable quantum parameters (Training).

All models were configured to have the same number of layers (20) to ensure a fair comparison. The results show that both HQNN variants with trainable quantum layers outperform their frozen counterparts and the classical NN baseline. Notably, NN-Angle (Training) achieved the lowest MAE ($1.082 \pm 0.004$), demonstrating the effectiveness of learning quantum parameters. Even in the freezing settings, both quantum-enhanced models achieved better or comparable performance to the classical NN, indicating that the QNN contributes to representation power even without being trained. This outcome aligns with the findings reported in~\cite{henderson2020quanvolutional}.

In addition, the QNN components in both NN-Amplitude and NN-Angle variants require only 540 quantum parameters, while the classical NN baseline requires 1,215 parameters for comparable depth. This highlights the parameter efficiency and scaling advantage of HQNN. As the input dimension increases, this advantage becomes more pronounced, as discussed further in Section~\ref{Sec:dis}.

\subsection{Analysis of Binding Affinity Prediction}
\begin{table*}[h]
\caption{Comprehensive Comparison with Qubit, Quantum Parameter, Classical Parameter, Prediction Accuracy of the Proposed HQDeepDTAF and Classical Counterparts on Protein-Ligand Binding Affinity Prediction of the PDBbind Dataset}
    \centering
    \begin{adjustbox}{width=\textwidth,center}
    \begin{tabular}{ccccccccc}
    \toprule
        Models                                 & \# Qubit                   & \# C Param.    & \# Q Param.              & MAE              & RMSE            & CI             & SD                & R \\ \hline
        DeepDTAF~\cite{wang2021deepdtaf}          & - & 154114 & -  & 1.109 $\pm$ 0.041 & 1.404 $\pm$ 0.049 & 0.785 $\pm$ 0.010 & 1.390 $\pm$ 0.056 & 0.769 $\pm$ 0.022 \\ \midrule
        DeepDTA~\cite{ozturk2018deepdta}          & - & 1523396 & -  & 1.208 $\pm$ 0.016 & 1.421 $\pm$ 0.019 & 0.775 $\pm$ 0.004 & 1.356 $\pm$ 0.019 & 0.740 $\pm$ 0.008 \\ \midrule
        Pafnucy~\cite{stepniewska2018development} & - & 15859081 & -  & 1.337 $\pm$ 0.015 & 1.573 $\pm$ 0.013 & 0.770 $\pm$ 0.004 & 1.559 $\pm$ 0.016 & 0.750 $\pm$ 0.007 \\ \midrule
        TopologyNet~\cite{cang2017topologynet}    & - & 33425345 & -  & 1.195 $\pm$ 0.032 & 1.406 $\pm$ 0.039 & 0.785 $\pm$ 0.011 & 1.392 $\pm$ 0.032 & 0.769 $\pm$ 0.024 \\ \midrule
        AEScore~\cite{aescore}                    & - & 155131 & -  & 1.359 $\pm$ 0.032 & 1.705 $\pm$ 0.036 & 0.726 $\pm$ 0.010 & 1.685 $\pm$ 0.042 & 0.632 $\pm$ 0.024 \\ \midrule
        Kdeep~\cite{kdeep}                        & - & 132954465 & -  & 1.182 $\pm$ 0.017 & 1.390 $\pm$ 0.022 & 0.786 $\pm$ 0.004 & 1.376 $\pm$ 0.021 & 0.771 $\pm$ 0.009 \\ \midrule
        \multirow{2}{*}{\makecell{HQDeepDTAF-\\Amplitude}} & \multirow{2}{*}{9} & \multirow{2}{*}{96551} & \multirow{2}{*}{540}  & \multirow{2}{*}{1.298 $\pm$ 0.084} & \multirow{2}{*}{1.631$ \pm$ 0.110} & \multirow{2}{*}{0.773 $ \pm $ 0.006} & \multirow{2}{*}{1.449 $ \pm $ 0.031} & \multirow{2}{*}{0.746 $ \pm $ 0.013} \\ \\ \midrule
        \multirow{2}{*}{\makecell{HQDeepDTAF-\\NN-Amplitude}} & 7 & 145829 & 420   & 1.104 $ \pm $ 0.042 & 1.388 $ \pm $ 0.056 & 0.791 $ \pm $ 0.010 & 1.373 $ \pm $ 0.058 & 0.776 $ \pm $ 0.024 \\
        & 9 & 145831 & 540  & 1.086 $ \pm $ 0.039 & 1.369 $ \pm $ 0.033 & 0.794 $ \pm $ 0.007 & 1.366 $ \pm $ 0.032 & 0.779 $ \pm $ 0.013 \\ \midrule
        \multirow{2}{*}{\makecell{HQDeepDTAF-\\NN-Angle}} & 7 & 99244 & 420      & 1.095 $ \pm $ 0.013 & 1.380 $ \pm $ 0.016 & 0.790 $ \pm $ 0.003 & 1.373 $ \pm $ 0.020 & 0.776 $ \pm $ 0.007 \\
                                                 & 9 & 100016 & 540   & 1.082 $ \pm $ 0.004 & 1.368 $ \pm $ 0.006 & 0.792 $ \pm $ 0.001 & 1.355 $ \pm $ 0.008 & 0.783 $ \pm $ 0.003 \\ 
    \botrule
    \end{tabular}
    \end{adjustbox}
\label{tab:comprehensive}
\end{table*}

To evaluate the effectiveness of the proposed HQDeepDTAF models in predicting protein-ligand binding affinity, we conduct a comprehensive comparison with classical baselines in terms of prediction accuracy and model complexity, as summarized in Table~\ref{tab:comprehensive}. The evaluation metrics include MAE, RMSE, $R$, SD, and CI, along with the number of classical and quantum parameters. 

Among the classical models, DeepDTAF achieves the best overall performance, recording the lowest MAE ($1.109$) and a competitive RMSE ($1.404$), while using only 0.15M classical parameters. This significantly outperforms other baselines such as DeepDTA (MAE$=1.208$, 1.52M parameters), Pafnucy (MAE$=1.337$, 15.8M), TopologyNet (MAE$=1.195$, 33.4M), and AEScore (MAE$=1.359$). These results highlight DeepDTAF's architectural efficiency, achieving high accuracy with significantly fewer parameters.
The proposed HQDeepDTAF models further explore hybrid quantum-classical architectures. Notably, the HQDeepDTAF variants (NN-Amplitude and NN-Angle) that incorporate a classical embedding network and quantum embedding consistently outperform the pure amplitude-based HQDeepDTAF across all accuracy metrics. Specifically, the HQDeepDTAF-NN-Angle model with 9 qubits achieves the best performance, yielding an MAE of $1.082$, RMSE of $1.368$, and the highest correlation coefficient ($R = 0.783$), along with the lowest prediction variance (SD$=1.355$).

Compared to other hybrid variants and classical baselines, the HQDeepDTAF-NN-Angle model demonstrates not only improved accuracy but also favorable parameter efficiency. It uses fewer classical parameters (100K) than HQDeepDTAF-NN-Amplitude (145K) and classical models like DeepDTA and Pafnucy, suggesting reduced training and inference costs. These results support the effectiveness of combining classical feature processing with quantum circuits, especially through data re-uploading and angle encoding, in achieving accurate and efficient binding affinity prediction under NISQ constraints.

\subsection{analysis of convergence speed}
\begin{figure}[h]
    \centering
    \includegraphics[width=0.6\linewidth]{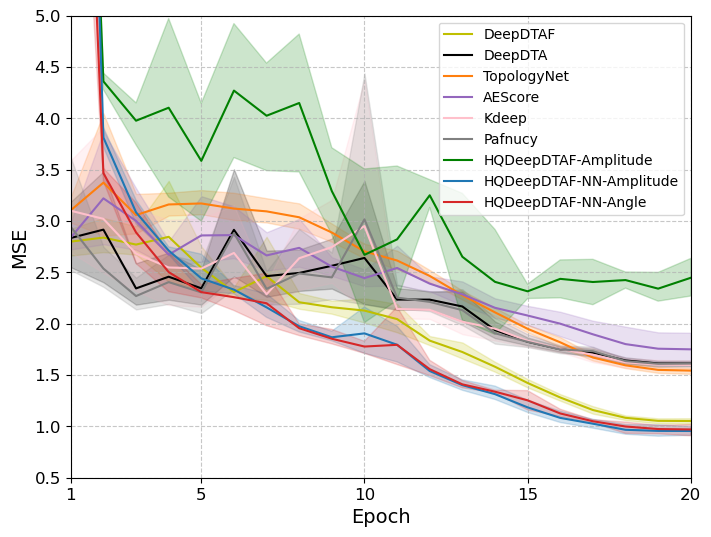}
    \caption{Convergence results of affinity prediction models}
    \label{fig:converge}
\end{figure}

To assess the training efficiency of each model, we further analyze the convergence behavior as illustrated in Fig.~\ref{fig:converge}. Most models converge within 20 epochs, demonstrating stable training dynamics. Notably, HQDeepDTAF-Amplitude and AEScore exhibit larger variance during training, indicating higher sensitivity to initialization or learning conditions. In addition, despite reproducing the original architecture and training settings, both AEScore and Pafnucy show signs of overfitting in our experiments.
Although the HQDeepDTAF-Amplitude model exhibits higher loss variance across runs, it achieves the fastest convergence speed among all models. However, when classical networks are integrated into the quantum model, such as in HQDeepDTAF-NN-Amplitude and HQDeepDTAF-NN-Angle, the convergence speed does not show a clear advantage over classical baselines in terms of the number of epochs. This suggests that incorporating hybrid quantum-classical structures does not necessarily result in faster convergence. This observation is similar with the findings reported in~\cite{moon2025qsegrnn}.
Nevertheless, this does not imply that quantum models lack training efficiency. It is important to note that all HQDeepDTAF variants use significantly fewer trainable parameters compared to their classical counterparts. As such, they require less computational cost for parameter updates, making them more efficient in terms of training resource consumption. These results suggest that HQDeepDTAF models are not only accurate but also parameter-efficient.

\subsection{Analysis of Noise Effects}\label{sec:noiseany}
\begin{table*}[ht]
\centering
\caption{Prediction performance (MAE) of HQDeepDTAF models and classical counterpart under depolarizing and amplitude damping noise with varying noise rate}
\label{tab:noise_comparison}
\begin{adjustbox}{width=\textwidth,center}
\begin{tabular}{lcccccccccc}
\toprule
\multirow{2}{*}{Models} & \multicolumn{5}{c}{Depolarizing Noise in Model}                  & \multicolumn{5}{c}{Amplitude Damping Noise in Model} \\ \cmidrule(lr){2-6} \cmidrule(lr){7-11}
                        & Noiseless & 0.001 & 0.01 & 0.1 & 0.2 & Noiseless & 0.001 & 0.01 & 0.1 & 0.2 \\ \midrule
DeepDTAF                & 1.109 & -- & -- & -- & -- & 1.109 & -- & -- & -- & -- \\
HQDeepDTAF-Amplitude    & 1.298 & 1.325 & 1.347 & 1.424 & 1.489 & 1.298 & 1.306 & 1.335 & 1.433 & 1.459 \\
HQDeepDTAF-NN-Amplitude & 1.086 & 1.096 & 1.108 & 1.136 & 1.140 & 1.086 & 1.110 & 1.144 & 1.140 & 1.153 \\
HQDeepDTAF-NN-Angle     & 1.082 & 1.090 & 1.096 & 1.126 & 1.133 & 1.082 & 1.083 & 1.099 & 1.118 & 1.122 \\
\bottomrule
\end{tabular}
\end{adjustbox}
\end{table*}

To develop the quantum algorithms considering NISQ devices, it is essential to take into account the impact of noise. In these devices, noise affects quantum states during unitary operations and measurements. In our experiments, we evaluate robustness to noise using two widely adopted noise models: depolarizing noise and amplitude damping noise a widely adopted method for noise simulation~\cite{moon2025qsegrnn, 10613453}. We simulate noise at rates of 0.001, 0.01, 0.1, and 0.2. All HQDeepDTAF models use 9 qubits with 20 layers.
For the proposed models, we set the number of qubits as $9$. 

Table~\ref{tab:noise_comparison} summarizes the prediction performance (MAE) under varying noise conditions. In the noiseless setting, the HQDeepDTAF-Hybrid models (NN-Amplitude and NN-Angle) outperform the classical baseline DeepDTAF. As noise is introduced, a gradual degradation in prediction accuracy is observed across all HQDeepDTAF models. This is due to the impact of noise on VQC gradient calculations, which is further explained in Appendix~\ref{secAppenD}. Under both noise types, the HQDeepDTAF-NN-Angle model consistently demonstrates the highest robustness, maintaining relatively low MAE values even as the error rate increases. At error rates of 0.001 and 0.01, all hybrid models (NN-Amplitude and NN-Angle) still outperform DeepDTAF. However, when the error rate reaches 0.1 or higher, all HQDeepDTAF models are affected significantly, and their performance drops below that of DeepDTAF.

These results demonstrate that HQDeepDTAF-Hybrid models perform better than DeepDTAF when no noise is present. Nonetheless, the superior performance of hybrid models in low-noise settings demonstrates their potential under realistic noise constraints. It is therefore crucial to assess quantum model behavior under expected noise levels to ensure practical viability on near-term hardware.

\section{Discussion}
\label{Sec:dis}
In this section, we discuss the efficiency and feasibility of the proposed model, along with practical considerations for implementation on NISQ devices. The efficiency and feasibility analyses are conducted under a theoretical framework~\cite{nielsen2010quantum}. To complement this, we also provide additional discussions addressing key practical challenges and design considerations relevant to NISQ-era quantum computing~\cite{bharti2022noisy}.

The efficiency of the proposed HQNN model was assessed under two conditions: 1) equal number of layers and 2) equal number of trainable parameters, compared to a classical NN. Under the same number of layers condition, we compared the number of trainable parameters, operation time, and model accuracy. Under the same number of parameter conditions, the comparison was conducted in terms of the number of layers, operation time, and accuracy. In particular, to quantify the cost of a quantum algorithm, the total number of gates and the circuit depth are useful~\cite{nielsen2010quantum}. For instance, $N$ gates can be processed in parallel. However, if $N$ gates must be executed in sequence, $N$ time steps are required. For the accuracy comparison, we evaluate the performance of our model against its classical counterpart using ablation studies.
Note that our analysis for efficiency focuses solely on HQNN and NN. This is because DeepDTAF and our proposed model share an identical CNN architecture.
To evaluate the model's feasibility on NISQ devices, we evaluated its complexity by considering the quantity of qubits, layers, gates, and parameters, where NISQ devices are considered IBM Quantum's quantum computers~\cite{abughanem2024ibmquantumcomputersevolution} for simplicity. 
We focused solely on the hybrid quantum aspects of the model, disregarding the embedding layer and classical convolution modules. Regarding the number of gates, we discuss the complexity of a circuit by the number of $\text{U}(4)$ gates since in many architectures they are more difficult to implement than $\text{U}(2)$ gates~\cite{malvetti2021quantum}. 

\paragraph{Efficiency of HQDeepDTAF}
\begin{figure}[ht]
\subfigure[Architecture of HQNN]{{\includegraphics[width=0.5\textwidth]{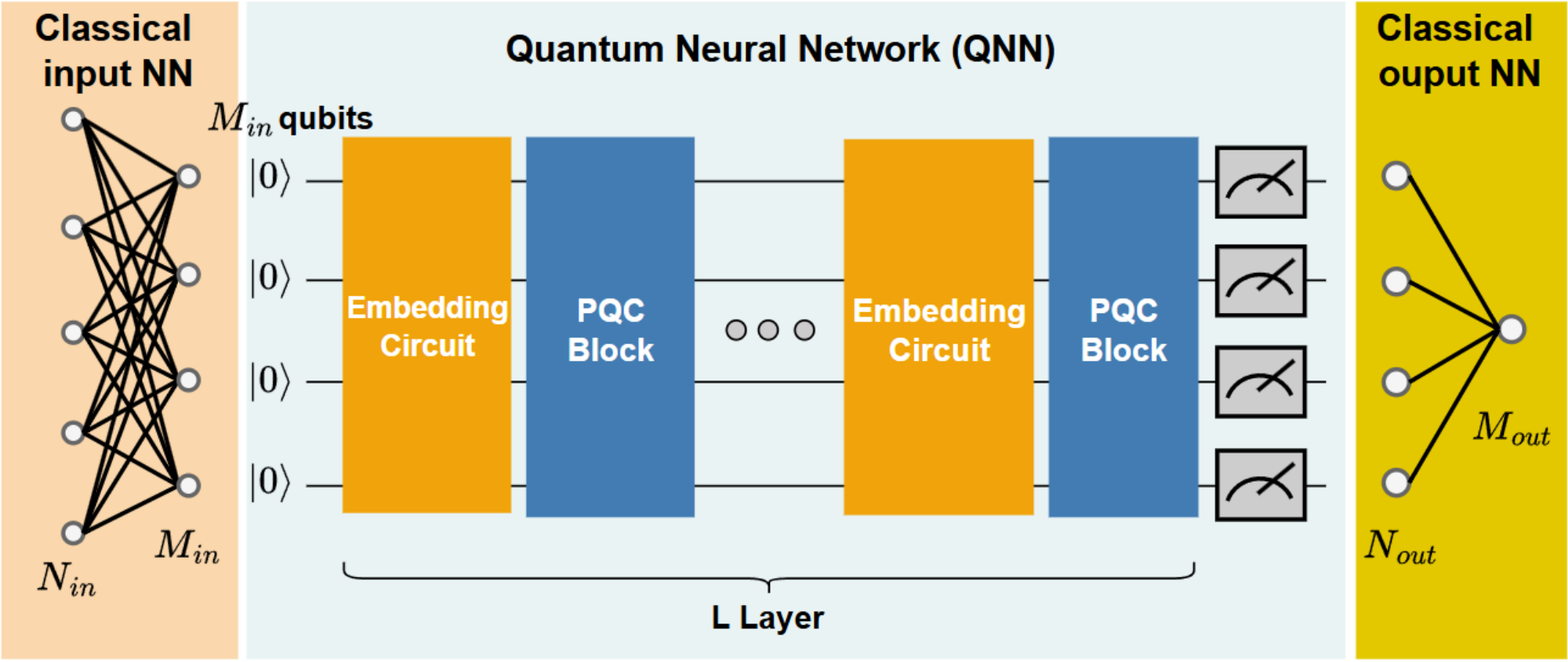} }}%
\label{fig.cq1}
\vspace{0cm}
\subfigure[Architecture of NN]{{\includegraphics[width=0.5\textwidth]{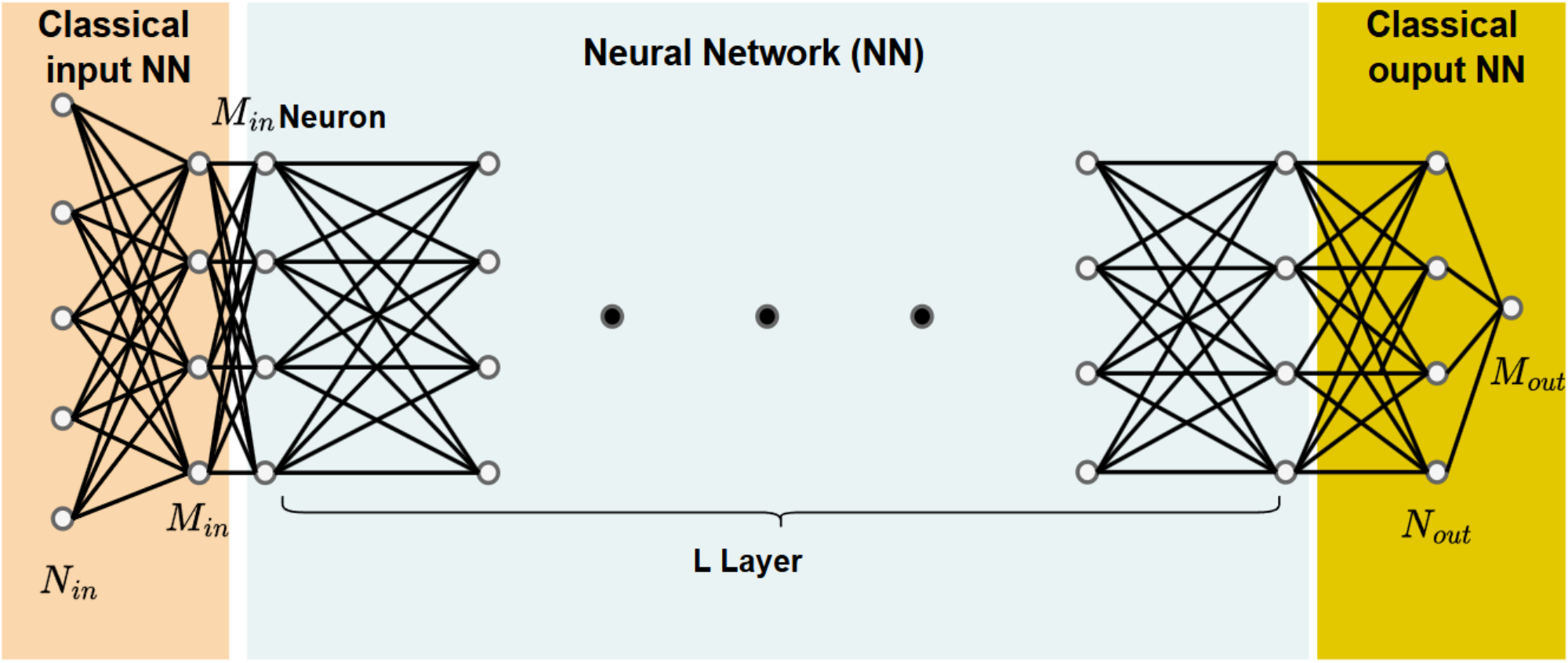} }}%
\label{fig.ct}
\caption{Architecture of HQNN and NN for efficiency analysis.}%
\label{fig.architecture}%
\end{figure} 

As shown in Fig.~\ref{fig.architecture}, in the efficiency analysis, for the generality, we consider the classical embedding network and classical regression network as classical input NN and output NN, respectively. That is, the HQNN consists of a classical input NN, QNN, and classical output NN sequentially in Fig.~\ref{fig.architecture}(a). For the classical NN, the classical NN comprises a classical input NN, NN, and classical output NN sequentially in Fig.~\ref{fig.architecture}(b). 

\begin{table*}[t]
\centering
\caption{Evaluation of efficiency of HQNN and NN under the same number of layers}
\label{Efficiency1}
\begin{adjustbox}{width=\textwidth,center} 
\begin{tabular}{llllll}
\toprule
Model    & \# Qubit   & \# C Params.                            & \# Q Params.       & \# Operation Time   \\  \midrule
Classical NN       & $-$        &$(N_{in}+M_{out}) M_{in}+M_{in}^2 (L+2)$ &  $-$              & $O((N_{in}+M_{out}) M_{in}+M_{in}^2 (L+2))\text{CPU}_{time}$                                 \\ \midrule
HQNN     & $M_{in}$   & $(N_{in}+M_{out})M_{in}$   & $3M_{in}L$        & $O((N_{in}+M_{out}) M_{in})  \text{CPU}_{time}+O(4L)\text{QPU}_{single}+O(M_{in} L) \text{QPU}_{two}$                \\ \botrule
\end{tabular}
\end{adjustbox}
\end{table*}

\begin{table*}[t]
\centering
\caption{Evaluation of efficiency of HQNN and NN under the same number of parameters}
\label{Efficiency2}
\begin{adjustbox}{width=\textwidth,center} 
\begin{tabular}{llllll}
\toprule
Model    & \# Qubit   & \# C Params.                            & \# Q Params.       & \# Operation Time   \\  \midrule
Classical NN       & $-$        &$(N_{in}+M_{out}) M_{in}+M_{in}^2 (L_{nn}+2)$ &  $-$              & $O((N_{in}+M_{out}) M_{in}+M_{in}^2 (L_{nn}+2))\text{CPU}_{time}$                                 \\ \midrule
HQNN     & $M_{in}$   & $(N_{in}+M_{out})M_{in}$   & $3M_{in}L_{qnn}$        & $O((N_{in}+M_{out}) M_{in})  \text{CPU}_{time}+O(4L_{qnn})\text{QPU}_{single}+O(M_{in} L_{qnn}) \text{QPU}_{two}$                \\ \botrule
\end{tabular}
\end{adjustbox}
\end{table*}

\textbf{Evaluation of the efficiency of HQNN and classical NN}: 
For the HQNN, Given $N_{in}$ input neuron of classical input NN, $M_{in}$ output neuron of classical input NN, $N_{out}$ input neuron of classical output NN, $M_{out}$ output neuron of classical output NN, the computational complexity of classical parts of HQNN can be calculated by $N_{in} M_{in}+N_{out} M_{out}$. Assume that all qubits are measured in QNN, then $N_{out}=M_{in}$. Therefore, the computational complexity can be rewritten as $(N_{in}+M_{out}) M_{in}$. For the QNN, we assume it is implemented as a data re-uploading QNN utilizing angle embedding. Given $M_{in}$-dimensional input to QNN, the required number of qubits is $O(M_{in})$. Due to angle embedding, the number of gate operations for encoding is $O(1)$ for each layer. The circuit depth of the embedding circuit is $O(L)$, where $L$ is the number of layers of QNN. With respect to the execution of quantum encoding circuits, the computational complexity is $O(L)$. For PQC blocks with the number of circuit layers $L$, the strongly entangling circuits have $3 M_{in} L~\text{U}(2)$ quantum gates and $M_{in} L~\text{U}(4)$ quantum gates. The required number of parameters for PQC blocks can be calculated as $3 M_{in} L$. The circuit depth is $O((3+M_{in}) L)$. Therefore, the total circuit depth of the QNN is $O((4+M_{in}) L)$, which is the same as the total execution of quantum circuits of QNN. The total number of quantum parameters of HQNN can be obtained as $O(3 M_{in} L)$. The HQNN was implemented using a combination of classical computers and quantum computers. That is, the operation time can be computed as $O((N_{in}+M_{out}) M_{in})  \text{CPU}_{time}+O(4L)\text{QPU}_{single}+O(M_{in} L) \text{QPU}_{two}$, where $\text{CPU}_{time}$ is process time on classical computer, $\text{QPU}_{single}$ denotes the $U(2)$ gate operation time on quantum processing unit (QPU), and $\text{QPU}_{two}$ means the $U(4)$ gate operation time on QPU. From these analyses, we can verify that HQNN has a polynomial operation time for layers and qubits.

Consider a classical NN with $L$ layers, where the initial layer has $N_{in}$ input neurons and $M_{in}$ output neurons. Each subsequent hidden layer contains $M_{in}$ neurons, while the final layer has $M_{out}$ neurons. The computational complexity of this NN can be expressed as $O((N_{in}+M_{out}) M_{in}+M_{in}^2 (L+2))$. The operation time is calculated by multiplying the computational complexity with the CPU time.

In the scenario with the same number of parameters, the classical NN and HQNN are configured with different numbers of layers to enable a fair comparison without loss of generality, as shown in Table~\ref{Efficiency2}. Here, $L_{nn}$ denotes the number of layers in the NN, and $L_{qnn}$ represents the number of layers in the QNN.

\textbf{Comparison}: First, we calculated the number of parameters and operation time for the first comparison scenario, which is an equal number of layers. 
In comparing the number of parameters, the number of parameters in the classical NN is consistently larger, except when $L = 1$ and $M_{in} = 1$.
With respect to the operation time,  we can disregard the $O((N_{in}+M_{out}) M_{in})$ term when comparing operation times, as this component is shared. That is, in terms of operation time, HQNN and NN are expressed as $O(4 L)\text{QPU}_{single}+O(M_{in} L) \text{QPU}_{two}$ and $O(M_{in}^2 (L+2)) \text{CPU}_{time}$, respectively, for comparison. In general, $\text{QPU}_{two}>QPU_{single}$~\cite{malvetti2021quantum}, we approximate HQNN operation time as $O((4+M_{in}) L) \text{QPU}_{two}$ for simple comparison. 
To achieve faster execution, the QPU time must satisfy the following conditions:
\begin{align}
    M_{in}^2(L+2)\text{CPU}_{time} &\geq (M_{in}+4)L\text{QPU}_{two},\\
    \frac{{M_{in}^2}(L+2)}{(M_{in}+4)L}CPU_{time} &\geq \text{QPU}_{two},
\label{QPU}
\end{align}
The inequality indicates that the HQNN remains advantageous as $M_{in}$ increases, since the classical model's complexity grows faster than that of the HQNN.

For the comparison of the second scenario, to achieve the same total number of parameters, the number of layers of HQNN can be calculated as follows: 
\begin{align}
    3M_{in}L_{qnn}&=M_{in}^2(L_{nn}+2),\\
    L_{qnn}&=\frac{M_{in}(L_{nn}+2)}{3}
\label{Lqnn}
\end{align}
To ensure that the above equation, $M_{in}(L_{nn}+2)$ must be divisible by $3$. This condition is satisfied if either $L_{nn}\equiv 1 (\text{mod}3)$ or $M_{in}\equiv 0(\text{mod 3})$. This constraint allows for deeper QNN architectures, thereby enhancing the model’s expressivity~\cite{perez2020data}.
In terms of operation time, and to ensure a fair comparison without loss of generality, the operation time of the HQNN is approximated as $O((4 + M_{in}) L_{qnn}) \text{QPU}_{two}$. To achieve faster execution, the QPU time must satisfy the following conditions:
\begin{align}
    M_{in}^2(L_{nn}+2)\text{CPU}_{time}&\geq (M_{in}+4)L_{qnn}\text{QPU}_{two},\\
    M_{in}^2(L_{nn}+2)\text{CPU}_{time}&\geq(M_{in}+4)\frac{M_{in}(L_{nn}+2)}{3}\text{QPU}_{two},\\
    \frac{3M_{in}}{M_{in}+4}\text{CPU}_{time}&\geq \text{QPU}_{two}.
\label{QPU}
\end{align}
It is noteworthy that the HQNN can outperform its classical counterpart in terms of operation time, even when the QPU is slower than the CPU. Specifically, the model remains advantageous if the QPU two-qubit gate time satisfies~\eqref{QPU}. This inequality implies that the QPU can be up to $3M_{in}/(M_{in}+4)$ times slower than the CPU, yet still achieve faster overall execution, owing to the structural efficiency of the HQNN architecture.

In summary, in the scenario with the same number of layers, the classical NN consistently requires more parameters. Moreover, the HQNN can achieve faster execution due to the classical NN's complexity grows faster than that of the HQNN.
In the scenario with the same number of parameters, the condition $M_{in}(L_{nn}+2) \equiv 0 \ (\text{mod }3)$ enables deeper QNNs and greater expressivity. In large-scale settings, HQNN can achieve faster operation time than classical NN due to the structural efficiency of the HQNN architecture.

\paragraph{Feasibility of HQDeepDTAF}
\begin{table*}[]
\centering
\caption{Evaluation of Complexity of HQDeepDTAF}
\label{complexity}
\begin{adjustbox}{width=\textwidth,center} 
\begin{tabular}{lllllll}
\toprule
Model      & \# Qubit & \# circuit depth & \# $\text{U}(4)$ gates & \# C Params. & \# Q Params \\  \midrule
HQDeepDTAF-NN-Amplitude & $M_{in}$  &   $O(poly(M_{in})L+(3+M_{in})L)$& $\frac{1}{4}(4^{M_{in}}-3M_{in}-1)L+M_{in}L$ &  $N_{in}2^{M_{in}}+M_{in}$ & $3M_{in}L$\\ \midrule
HQDeepDTAF-NN-Angle     & $M_{in}$   &  $O(4+M_{in})L)$  &  $M_{in}L$ & $(N_{in}+1)M_{in}$ & $3M_{in}L$\\ \botrule
\end{tabular}
\end{adjustbox}
\end{table*}
To evaluate the feasibility of the HQDeepDTAF for regression task on the NISQ device, we assume that the input dimension of the classical embedding network is $N_{in}$, and the number of qubits $M_{in}$. 

1) HQDeepDTAF-NN-Amplitude: Given a classical embedding network with $N_{in}$ input neurons and $2^{M_{in}}$ output neurons, the computational complexity is computed as $O(N_{in}2^{M_{in}})$. Given $2^{M_{in}}$-dimensional input to QNN, the required number of qubits is $\lceil \log(2^{M_{in}}) \rceil$. For simplicity, we assume that the required number of qubits is ${M_{in}}$. Considering $L$ layers and data re-uploading scheme, the circuit depth of the amplitude embedding circuit is $O(poly(M_{in})L)$. Note that the best known lower bound for number of CNOT gates for amplitude embedding is $\frac{1}{4}(4^{M_{in}}-3{M_{in}}-1)$~\cite{sun2023asymptotically}. Therefore, the total number of $\text{U}(4)$ gates is calculated as $\frac{1}{4}(4^{M_{in}}-3{M_{in}}-1)L+{M_{in}}L$, where the strongly entangling layer~\cite{schuld2020circuit} is used as PQC ansatz.
Finally, the total circuit depth of the QNN with amplitude embedding is $O(poly(M_{in})L+(3+M_{in})L)$. The total number of quantum parameters of QNN can be obtained as $O(3M_{in}L)$. The classical regression network is used for regression task. The output dimension of the classical regression network is $1$. Therefore, the computational complexity of the classical regression network can be calculated as $O(M_{in})$.

2) HQDeepDTAF-NN-Angle: In the classical embedding network, the number of output neurons corresponds to $O(M_{in})$, which matches the qubit count in the QNN. The computational complexity of this embedding network is calculated as $O(N_{in}M_{in})$. The circuit depth of the angle embedding circuit is $O(L)$. Therefore, the total circuit depth of the QNN with angle embedding is $O((4+{M_{in}})L)$. Finally, the total number of quantum parameters of QNN and the computational complexity of the classical regression network is the same as the HQDeepDTAF-NN-Amplitude. 

\textbf{Comparison}: Table~\ref{complexity} shows the complexity details of our models.
We can verify that the HQDeepDTAF-NN-Amplitude requires significantly long coherence time and polynomial $\text{U}(4)$ gates. This implies more difficult to implement in practice. In contrast, HQDeepDTAF-NN-Angle is more efficient and feasible in terms of circuit depth, $\text{U}(4)$ gates, classical parameter, and quantum parameter. 
In addition, we prove the feasibility of the HQDeepDTAF-NN-Angle on NISQ devices in the following. 
\begin{proof}
For a view of the availability of IBM Quantum's current quantum systems, the proposed models utilized $9$ qubits, which is achievable in current NISQ devices~\cite{abughanem2024ibmquantumcomputersevolution}.
With respect to the required time to execute the quantum circuits, we first determine the specific circuit depth using the experimental values. For the HQDeepDTAF-NN-Amplitude, a depth for embedding circuit is calculated upper bound $O(2^{M_{in}}L)$ reffering~\cite{sun2023asymptotically}. Therefore, given $9$ qubits and $20$ layers, the circuit depth of HQDeepDTAF-NN-Angle and HQDeepDTAF-NN-Amplitude are calculated as $260$ and $10480$, respectively. Currently, most IBM quantum computers support a maximum of $300$ circuits~\cite{abughanem2024ibmquantumcomputersevolution}. Therefore, the only HQDeepDTAF-NN-Angle model can be implemented in the current IBM quantum computer.  
\end{proof}

\paragraph{Practical Considerations for NISQ Implementation.}
While our analysis is closer to fault-tolerant models~\cite{nielsen2010quantum}, practical deployment on NISQ hardware requires additional care due to experimental limitations such as quantum noise, decoherence, restricted qubit connectivity, and native gate sets~\cite{bharti2022noisy}. Our evaluation does not explicitly incorporate qubit connectivity or quantum error mitigation (QEM) techniques, which we acknowledge as limitations of our current framework. Below, we discuss our model from the perspectives of noise robustness and practical feasibility on real NISQ hardware.

\begin{enumerate}
    \item \textbf{Noise Robustness:} Quantum noise is a central challenge in NISQ computing. To assess the robustness of our model, we conducted noise-aware experiments using hardware-realistic noise models. As shown in our results (Section~\ref{sec:noiseany}), the performance of HQDeepDTAF noticeably degrades under noisy conditions. This observation highlights the need for incorporating QEM techniques such as zero-noise extrapolation or probabilistic error cancellation~\cite{temme2017error} in future implementations. While QEM was not applied in this study, our HQDeepDTAF architecture is compatible with such extensions, and integrating them remains an important direction for improving robustness.

    \item \textbf{Practical Feasibility:} For NISQ feasibility, we demonstrated that the HQDeepDTAF-NN-Angle is more efficient and feasible in terms of circuit depth. While our analysis assumes universal gate operations, actual hardware supports only native gate sets. By the Solovay–Kitaev theorem~\cite{dawson2005solovaykitaevalgorithm}, any unitary operation can be approximated with a logarithmic-length gate sequence from a universal set, introducing additional overhead not captured in our theoretical model. Thus, we recognize that gate synthesis, connectivity constraints, and compilation must be considered in future hardware-specific implementations.
\end{enumerate}

\section{Conclusion}
This study proposed the HQDeepDTAF algorithm for protein-binding affinity prediction that can be implemented on NISQ devices. Given the limitations of current quantum devices, including qubit availability, noise, and short coherence time, a hybrid quantum model was developed. In addition, this study empirically explored the potential of HQNNs to serve as substitutes for classical NNs by evaluating their capacity for non-linear function approximation. Through ablation studies, we demonstrated that HQNNs effectively combine the representational strengths of classical and quantum components, highlighting their practical viability under NISQ constraints. Experimental results showed that HQDeepDTAF-NN-Angle and -Amplitude models outperformed benchmarks in protein-binding affinity prediction in terms of the number of parameters and prediction accuracy.  
To assess the algorithm's effectiveness, noise simulation using the depolarizing channel and amplitude damping channel was performed, confirming the model's resilience to noise. Nevertheless, it is important to note that excessive noise levels can lead to a decline in model effectiveness, making this a crucial factor to take into account. In the discussion, the study evaluated the efficiency and feasibility of the proposed model, comparing HQNN operation time to classical NN and demonstrating faster performance. Feasibility assessment revealed that only the HQDeepDTAF-NN-Angle model is currently implementable on IBM quantum computers, while the HQDeepDTAF-NN-Amplitude model is not feasible due to NISQ device limitations. Furthermore, while additional quantum layers of our model could enhance its performance, the limitations of current NISQ devices prevent this improvement. Although our study explores the potential for quantum advantage, it is important to note that all results are derived from classical simulations. We do not claim a definitive quantum advantage, and further investigation on real quantum devices is necessary to assess this possibility from various perspectives.
Thus, to demonstrate our findings, we will evaluate the HQDeepDTAF models in real quantum computers. Moreover, QEM strategies represent a critical direction for future research toward enhancing noise resilience and strengthening the feasibility of quantum models on NISQ devices.

\newpage
\begin{appendices}
\section{Abbreviations}
\label{secacronym}

\begin{table}[h]
\caption{Abbreviations}
\label{acrym}
\begin{tabular}{ll}
\toprule
Acronym  & Description  \\
\midrule
AI       & Artificial Intelligence \\
CNNs     & Convolutional Neural Networks \\
CPU      & Central Processing Unit \\
CI       & Concordance Index \\
FC       & Fully Connected \\
FRQI     & Flexible Representation of Quantum Image \\
HQDeepDTAF & Hybrid Quantum DeepDTAF \\
HQNN     & Hybrid Quantum Neural Network \\
KL       & Kullback-Leibler \\
MAE      & Mean Absolute Error \\
ML       & Machine Learning \\
MSE      & Mean Squared Error \\
NISQ     & Noisy Intermediate-Scale Quantum \\
NNs      & Neural Networks \\
PDBbind  & Protein Data Bank Bind \\
PQC      & Parameterized Quantum Circuit \\
QML      & Quantum Machine Learning \\
QNNs     & Quantum Neural Networks \\
QPU      & Quantum Processing Unit \\
R        & Pearson Correlation Coefficient \\
ReLU     & Rectified Linear Unit \\
RMSE     & Root Mean Square Error \\
SD       & Standard Deviation \\
SDF      & Structure Data Format \\
SMILES   & Simplified Molecular Input Line Entry System \\
SSEs     & Secondary Structure Elements \\
UAP      & Universal Approximation Property \\
UAT      & Universal Approximation Theorem \\
\bottomrule
\end{tabular}
\end{table}

\section{Decomposition of PQC Block}
\label{app1}
Since the single-qubit rotations $R$ is a \text{U}(2), the rows and columns of $R$ are orthonormal, from which it follows that there exist real numbers  $\alpha, \beta, \gamma,~\text{and}~\delta$ such that follows~\cite{barenco1995elementary}:
\begin{equation}
R(\alpha, \beta, \gamma, \delta)=e^{i\delta}
\begin{pmatrix}
e^{i\beta}\cos(\alpha) & e^{i\gamma}\sin(\alpha)\\
-e^{-i\gamma}\sin(\alpha) & e^{-i\beta}\cos(\alpha)
\end{pmatrix},
\label{Req}
\end{equation}
where $\delta$ denotes the overall phase factors, which is commonly neglected due to difficulty to physically measure. Therefore, the single-qubit rotations can be considered only three trainable parameters per gate, $R(\alpha, \beta, \gamma)$.

In this paper, we decompose the single-qubit rotations $R\in \textbf{SU}(2)$ into the product of three rotations~\cite{bergholm2022pennylaneautomaticdifferentiationhybrid}, which is obtained as follows:
\begin{equation}
\begin{split}
R(\alpha, \beta, \gamma)&=R_{z}(\gamma)R_{y}(\beta)R_{z}(\alpha)\\&=
\begin{pmatrix}
e^{-i(\frac{\alpha+\gamma}{2})}\cos(\frac{\beta}{2}) & -e^{i(\frac{\alpha-\gamma}{2})}\sin(\frac{\beta}{2})\\
e^{-i(\frac{\alpha-\gamma}{2})}\sin(\frac{\beta}{2}) & e^{i(\frac{\alpha+\gamma}{2})}\cos(\frac{\beta}{2})
\end{pmatrix}.
\end{split}
\label{Req1}
\end{equation}

\section{Noise Effect for Gradient}
\label{secAppenD}
For an given initial quantum state $\rho$ with a VQC $U(\theta)$, the expectation value is represented as follows:
\begin{equation}\notag
\langle \hat{B}(\theta)\rangle = \text{Tr}(\hat{B}U(\theta)\rho U^{\dagger}(\theta)),
\label{measurement1}
\end{equation}
where $\hat{B}$ is the measurement of a final observable $\hat{B}$.

For the optimization process, the gradients of VQCs are computed based on the parameter-shif rule as follows:
\begin{equation}\notag
\grad_{\theta} \langle \hat{B} \rangle(\theta) = \frac{1}{2}\left[\langle \hat{B} \rangle(\theta+\frac{\pi}{2})-\langle \hat{B} \rangle(\theta-\frac{\pi}{2})\right].
\label{measurement2}
\end{equation}
In the noisy channel $\mathcal{E}$, the expectation value applied noise on the VQCs is rewritten as follows:
\begin{equation}\notag
\begin{split}
\langle \hat{B}(\theta)\rangle &= \text{Tr}(\hat{B}\mathcal{E}\left[U(\theta)\rho U^{\dagger}(\theta)\right])\\
&=\text{Tr}(\hat{B}'\left[U(\theta)\rho U^{\dagger}(\theta)\right])\\
&=\langle \hat{B}' \rangle (\theta),
\label{measurement3}
\end{split}
\end{equation}
where $\mathcal{E}$ denotes the noisy channel. $\hat{B}'=\mathcal{E}^{\dagger}[\hat{B}]$ means a new observable. This impacts the parameter-shift rule, the gradient of the VQCs is rewritten as follows:
\begin{equation}\notag
\grad_{\theta} \langle \hat{B}' \rangle(\theta) = \frac{1}{2}\left[\langle \hat{B}' \rangle(\theta+\frac{\pi}{2})-\langle \hat{B}' \rangle(\theta-\frac{\pi}{2})\right].
\label{measurement3}
\end{equation}

\end{appendices}

\section{Declarations}

\subsection{Availability of data and materials}
Our code is available at \url{https://github.com/drmoon-1st/HQDeepDTAF}

\subsection{Competing interests}
The authors declare that they have no competing interests.

\subsection{Funding}
This work was supported by Quantum Computing based on Quantum Advantage challenge research(RS-2024-00408613) through the National Research Foundation of Korea(NRF) funded by the Korean government (Ministry of Science and ICT(MSIT)). We would like to thank for support from Korea Agency for Infrastructure Technology Advancement(KAIA) grant funded by the Ministry of Land Infrastructure and Transport (No. RS-2023-00256816), and from the National Research Foundation of Korea(NRF) grant funded by the Korea government(MSIT)(RS-2024-00336962). We awknowledge additional support from the Institute of Information and communications Technology Planning and Evaluation (IITP) under the Artificial Intelligence Convergence Innovation Human Resources Development (IITP-2025-RS-2023-00254177) grant funded by the Korea government(MSIT).

\subsection{Author contributions}
S.G.J. contributed to the methodology, implementation, and experiment, and prepared the manuscript. K.H.M. contributed to the experiment and prepared the manuscript. W.J.H. supervised the project. All authors reviewed the manuscript.

\subsection{Acknowledgements}
Not applicable.


\bibliography{sn-bibliography}

\end{document}